\documentclass[11pt,twocolumn]{aastex631}

\received{July 23, 2022}
\revised{August 27, 2022}
\accepted{August 30, 2022}
\submitjournal{ApJL}
\usepackage{ulem}

\newcommand{\pmra}{$\mu_\alpha \cos\delta$}
\newcommand{\pmdec}{$\mu_\delta$}

\newcommand{\kms}{$\rm km\,s^{-1}$ }

\newcommand{\kmsnospace}{$\rm km\,s^{-1}$}

\begin{document}

\title{Dynamical Origin for the Collinder~132-Gulliver~21 Stream: \\A Mixture of three Co-Moving Populations with an Age Difference of 250\,Myr
}
   
\shorttitle{The Collinder\,132-Gulliver\,21 stream}
\shortauthors{Pang et al.}

\author[0000-0003-3389-2263]{Xiaoying Pang}
    \affiliation{Department of Physics, Xi'an Jiaotong-Liverpool University, 111 Ren’ai Road, Dushu Lake Science and Education Innovation District, Suzhou 215123, Jiangsu Province, P. R. China}
    \email{Xiaoying.Pang@xjtlu.edu.cn}
    \affiliation{Shanghai Key Laboratory for Astrophysics, Shanghai Normal University, 
                100 Guilin Road, Shanghai 200234, P. R. China}
    
\author{Yuqian Li}
    \affiliation{Department of Physics, Xi'an Jiaotong-Liverpool University, 111 Ren’ai Road, Dushu Lake Science and Education Innovation District, Suzhou 215123, Jiangsu Province, P. R. China}
    
\author[0000-0003-4247-1401]{Shih-Yun Tang}
    \affiliation{Lowell Observatory, 1400 W. Mars Hill Road, Flagstaff, AZ 86001, USA}
    \affiliation{Department of Astronomy and Planetary Science, Northern Arizona University, Flagstaff, AZ 86011, USA}

\author[0000-0001-8713-0366]{Long Wang}
    \affiliation{School of Physics and Astronomy, Sun Yat-sen University, Daxue Road, Zhuhai, 519082, China}
    \affiliation{CSST Science Center for the Guangdong-Hong Kong-Macau Greater Bay Area, Zhuhai, 519082, China}

\author[0000-0002-1243-8224]{Yanshu Wang}
    \affiliation{Department of Physics, Xi'an Jiaotong-Liverpool University, 111 Ren’ai Road, Dushu Lake Science and Education Innovation District, Suzhou 215123, Jiangsu Province, P. R. China}
    
\author[0000-0001-5017-7021]{Zhaoyu Li}
    \affiliation{Department of Astronomy, School of Physics and Astronomy, Shanghai Jiao Tong University, 800 Dongchuan Road, Shanghai 200240, P. R. China}

\author{Danchen Wang}
    \affiliation{Department of Physics, Xi'an Jiaotong-Liverpool University, 111 Ren’ai Road, Dushu Lake Science and Education Innovation District, Suzhou 215123, Jiangsu Province, P. R. China}

\author[0000-0002-1805-0570]{M.B.N. Kouwenhoven}
    \affiliation{Department of Physics, Xi'an Jiaotong-Liverpool University, 111 Ren’ai Road, Dushu Lake Science and Education Innovation District, Suzhou 215123, Jiangsu Province, P. R. China}

\author[0000-0003-3784-5245]{Mario Pasquato}
    \affiliation{Center for Astro, Particle and Planetary Physics (CAP$^3$), New York University Abu Dhabi, United Arab Emirates}
    \affiliation{INFN- Sezione di Padova, Via Marzolo 8, I–35131 Padova, Italy}

\begin{abstract} 

We use Gaia DR\,3 data to study the Collinder\,132-Gulliver\,21 region via the machine learning algorithm \textsc{StarGO}, and find eight subgroups of stars (ASCC\,32, Collinder\,132\,gp\,1--6, Gulliver\,21) located in close proximity. Three co-moving populations were identified among these eight subgroups: (i) a coeval 25\,Myr-old moving group (Collinder\,132); (ii) an intermediate-age (50--100\,Myr) group; and (iii) the 275\,Myr-old dissolving cluster Gulliver\,21. These three populations form parallel diagonal stripe-shape over-densities in the $U$--$V$ distribution, which differ from open clusters and stellar groups in the solar neighborhood. We name this kinematic structure the {\em Collinder\,132-Gulliver\,21 stream}, as it extends over 270\,pc in the 3D space. The oldest population Gulliver\,21 is spatially surrounded by  the Collinder\,132 moving group and the intermediate-age group. Stars in the Collinder\,132-Gulliver\,21 stream have an age difference up to 250\,Myr. Metallicity information shows a variation of 0.3~dex between the youngest and oldest populations. The formation of the Collinder\,132-Gulliver\,21 stream involves both star formation and dynamical heating. The youngest population (Collinder\,132 moving group) with homogeneous metallicity is probably formed through filamentary star formation. The intermediate-age and the oldest population were then scatted by the Galactic bar or spiral structure resonance to intercept Collinder\,132's orbit. Without mutual interaction between each population, the three populations are flying by each other currently and will become distinct three groups again in approximately $\sim$50\,Myr.

\end{abstract}

\keywords{stars: evolution --- open clusters and associations: individual -- stars: kinematics and dynamics -- methods: statistical -- methods: numerical }


\section{Introduction}

Moving groups are congregations of co-moving stars that typically extend from a few hundred parsecs to a few kilo-parsec in space. Most moving groups known in the solar neighborhood \citep[e.g.,][]{eggen1996,skuljan_velocity_1999} are named after open clusters in the region, such as the Hyades, the Pleiades, Coma Berenices, and IC\,2391 moving groups. Some of these moving groups are coeval and are thought to have originated from dissolving open clusters \citep[e.g.,][]{miret-roig_dynamical_2020,gagne_number_2021,lee_low-mass_2022,messina_gyrochronological_2022}. Other young coeval moving groups, on the other hand, were thought to form simultaneously in the same giant molecular cloud (GMC) \citep{kounkel2019}.

Most of the moving groups, however, are not associated with their eponymous open clusters and appear to host populations of different ages and inhomogeneous metallicities, such as the Hercules, Arcturus, and HR\,1614 moving groups \citep[][]{eggen1996,bensby_disentangling_2007,bensby_exploring_2014,kushniruk_hr_2020}. An age spread of several billion years found in the aforementioned moving groups \citep{kushniruk_hr_2020} cannot be simply explained by inhomogeneous star formation.

Stars in moving groups share similar kinematics like stars in stellar clusters; however, moving groups stand out in the velocity distribution as elongated substructures \citep[e.g., horizontal or diagonal branches on the $U$--$V$ velocity plane,][]{antoja_origin_2008,zhao2009,kushniruk2017,gaia2018kin,kushniruk_hr_2020} oppose to a concentrated distribution of stellar clusters. A horizontal arch shown on the $U$--$V$ velocity space can indicate conservation of vertical angular momentum, which might  support a dynamical origin of the moving group \citep{kushniruk_hr_2020}.

Various theories have sprung up to provide solutions to the formation mechanism of non-coeval moving groups. \citet{s_de_simone_stellar_2004} proposed that stochastic transient spiral waves can heat up the disk and generate moving groups, which mainly excite the stellar horizontal velocity components along the Galactic disk. Other formation theories are mostly related to resonances. The Sirius and Hyades moving groups have been suggested to be induced by the inner Lindblad resonance of the spiral structures. On the other hand, the Hercules moving group is associated with the Galactic bar's outer Lindblad resonance \citep{dehnen_effect_2000,bovy_velocity_2010}. The Coma Berenices and Pleiades moving groups may be related to the spiral co-rotation resonance \citep{barros_exploring_2020}. 
Another alternative is an external perturbation triggered by a minor merger event \citep{minchev_low-velocity_2010,antoja_kinematic_2012,barros_exploring_2020}, which produces vertical phase mixing  \citep{antoja_dynamically_2018,li2020,li2021} and excites the stars to move perpendicular to the Galactic disk. When stars in a moving group are formed at different locations and/or at different times, they will have different ages and come with different metallicities.

The young moving group studied in this paper, Collinder\,132, was originally known as an open cluster by \citet{collinder1931} and later suggested to host two populations by \citet{claria1977,eggen1983}. \citet{kounkel2019} later used Gaia DR\,2 data and suggested Collinder\,132 to be a coeval moving group extending 197\,pc with an age of 25\,Myr. Meanwhile, in the same sky region as Collinder\,132, a cluster, Gulliver\,21, with an age ten times older than Collinder\,132 was found \citep[275\,Myr,][]{cantat2018,pang2022}.
As no investigation has been made for Collinder\,132 and Gulliver\,21 yet, whether or not this older cluster, Gulliver\,21, is associated with Collinder\,132 moving group is still unknown. If an association exists, the origin of this moving group can be further constrained by the dynamical formation mechanism.

The latest Gaia Data Release \citep[DR\,3,][]{gaia_collaboration_gaia_2022} published radial velocity (RV) measurements for 34 million stars, a dataset four times larger than the Gaia Data Release~2 \citep[DR\,2,][]{gaia2018}. We aim to study the possible connection between the Collinder\,132 moving group and the cluster Gulliver\,21 using Gaia DR\,3 kinematic data. A kinematic relationship between the 275\,Myr-old cluster Gulliver\,21, and the 25\,Myr-old moving group Collinder\,132 will provide strong observational constraints to the formation and evolution of moving groups.

This paper is organized as follows. In Section~\ref{sec:gaia} we discuss the quality and limitations of the Gaia DR\,3 and EDR\,3 data. We then present the algorithm, \textsc{StarGO}, which is used to determine members in groups in Section~\ref{sec:stargo}. In Section~\ref{sec:multi}, we identify three moving groups in kinematic space and name the kinematic structures made of these three populations the Collinder\,132-Gulliver\,21 stream. The spatial characteristics and dynamical state of the stream are presented in Section~\ref{sec:stream}. The origin of the Collinder\,132-Gulliver\,21 stream is investigated in Section~\ref{sec:origin}, in which we 
discuss the possibility of different dynamical formation mechanism. Finally, we provide a brief summary of our findings in Section~\ref{sec:summary}.

\section{Data and Membership Identification}\label{sec:data_member}
\subsection{Gaia EDR3 and DR3 Data}\label{sec:gaia}

The Gaia DR\,3 \citep{gaia_collaboration_gaia_2022} became publicly available 1.5 years after the release of EDR\,3 \citep[released on December 3, 2020][]{gaia2021}. Parts of the data in DR3 are inherited from EDR3, such as the full astrometric solution (e.g., sky positions, parallaxes, and proper motions) and G$_{\rm BP}$ and G$_{\rm RP}$ magnitudes. Correction to the G-band photometry has been made in DR3 \citep{riello2021}, which mainly affects sources fainter than G=13\,mag. In the sky region of Collinder\,132 and Gulliver\,21, the mean difference in G-band magnitude between EDR3 and DR3 is approximately 0.003~mag, and reaches a maximum value of 0.025\,mag. 

The number of targets with RV measurement from the Radial Velocity Spectrometer (RVS, with a median resolving power R$\sim$11500) increased from $\sim$7 million in DR2 to $\sim$34 million in DR3. The longer observation baseline, 34 months of the nominal mission, also helped push the processing limitation from $G_{\rm RVS}$=12\,mag in Gaia DR2, to $G_{\rm RVS}$=14\,mag in DR3. The median precision of the RV is 1.3\,\kms at $G_{\rm RVS}$ = 12\,mag and 6.4\,\kms at $G_{\rm RVS}$=14\,mag \citep{katz_gaia_2022}.

With the release of Gaia DR3, astrophysical parameters have become available; these are either derived from RVS spectra and/or from low-resolution (R$\sim$40) BP/RP prism spectra \citep[][]{creevey_gaia_2022}. 
Astrophysical parameters determined from forward-modeling the BP/RP spectra  \citep[GSP-Phot, 470 million;][]{andrae_gaia_2022} outnumber those obtained from combined RVS spectra of single stars  \citep[GSP-Spec,6 million;][]{recio-blanco_gaia_2022}, which comes with more precise and detailed information on individual chemical abundances.

\subsection{Member Determination}\label{sec:stargo}

The member identification process for the stellar groups discussed in this paper is carried out as described in \citet{tang2019} and \citet{pang2020,pang2021a,pang2021b,pang2022}. In short, the membership identification process is performed in a sequence of three steps. First, we make two spherical cut in the 3D Cartesian coordinates space with a radius of 150~pc from the center of Collinder\,132 obtained from \citet{Cantat-Gaudin2020}, and a radius of 150~pc from the center of ASCC\,32 taken from \citet{liu2019}, in order to cover the elongated morphology of Collinder\,132 moving group. Second, we perform a further proper motion (PM) cut on the target region based on a 2D density map \citep[e.g., figure~1 in][]{pang2021a}. These circular PM cuts are performed to include as many potential members as possible and to reduce the number of field stars that can weaken the clustering signature in the member identification process. Third, we use the unsupervised machine-learning method \textsc{StarGO} \citep{yuan2018}\footnote{\url{https://github.com/zyuan-astro/StarGO-OC}} to identify the grouping using a 5D data set, i.e., $X$, $Y$, $Z$, \pmra, and \pmdec. The \textsc{StarGO} software is based on the Self-Organizing Map (SOM) algorithm, which can help mapping high dimensional data onto a 2D neural network (Figure~\ref{fig:som}) and searching for grouping. We provide more details on member selection with \textsc{StarGO} in Appendix~\ref{sec:som}.

\textsc{StarGO} has proven to be successful in the identification of  stellar streams and star cluster membership  \citep{tang2019,Zhang2019,yuan2020a,yuan2020b,pang2020,pang2021a,pang2021b,pang2022}. The algorithm \textsc{StarGO} is not only efficient in mapping the detailed structure of each stellar cluster or group, but is also able to identify hierarchical structures in stellar groups. \citet{pang2021b} use this top-down identification approach to disentangle five second-level substructures in the Vela\,OB2 region and \citet{pang2022} identified ten new hierarchical groups in four young regions using \textsc{StarGO}.

In total, we identify eight subgroups of stars in the  Collinder\,132-Gulliver\,21 region. Among these, Collinder\,132\,gp\,1 (pink patch in Figure~\ref{fig:som} (b)) and Collinder\,132\,gp\,2 (orange patch) are second-level structures on the SOM. Collinder\,132\,gp\,1 (corresponding to the stellar group Collinder\,132), Gulliver\,21 (cyan patch) and ASCC\,32 (blue patch) are known open clusters or stellar groups that were reported in previous catalogs \citep{liu2019,Cantat-Gaudin2020,he_blind_2022}. The remaining five groups are new, and are labelled as Collinder\,132\,gp\,2--6.



\begin{deluxetable*}{lCC C CCCR C LL RR RR R}
\tablecaption{General properties of eight subgroups in the Collinder\,132-Gulliver\,21 stream \label{tab:general}
             }
\tabletypesize{\scriptsize}
\tablehead{
	 \colhead{Cluster}  &
	 \colhead{R.A.}     & \colhead{Decl.}   &&
	 \colhead{Dist.}    &
	 \colhead{$X_c$}    & \colhead{$Y_c$}   & \colhead{$Z_c$}   & 
	 \colhead{RV}       &
     \colhead{\pmra}    & \colhead{\pmdec}  &
     \colhead{Age}      & 
	 \colhead{$M_{cl}$} & 
	 \colhead{$r_{\rm h}$} & \colhead{$r_{\rm t}$} & 
	 \colhead{$N$}      \\
	 \colhead{}                 & 
	 \multicolumn{2}{c}{(deg)}  &&
	 \multicolumn{4}{c}{(pc)}   &
	 \colhead{(km~s$^{-1}$)}    &
	 \multicolumn{2}{c}{(mas yr$^{-1}$)}    &
	 \colhead{(Myr)}            &
	 \colhead{($M_\odot$)}      & 
	 \multicolumn{2}{c}{(pc)}   & 
	 \colhead{}                 \\
	 \cline{2-3} \cline{5-8}   \cline{9-11} \cline{14-15}
	 \colhead{(1)} & \colhead{(2)} & \colhead{(3)} && \colhead{(4)} & \colhead{(5)} &
	 \colhead{(6)} & \colhead{(7)} & \colhead{(8)} & \colhead{(9)} & \colhead{(10)} &
	 \colhead{(11)} & \colhead{(12)} & \colhead{(13)} & \colhead{(14)} &
	 \colhead{(15)} 
	 }
\startdata
\rm ASCC\,32	&	105.730278	&	-26.449193	&&	795.2	&	-416.7	&	-664.6	&	-130.3	&	32.4	 &	-3.228	&	3.475	&	25	&	577.6	&	17.4	&	11.7	&	519	\\
\rm Collinder\,132\,gp\,1	&	108.870127	&	-31.069356	&&	655.1	&	-287.8	&	-579.4	&	-103.2	&	21.7  	&	-4.227	&	3.756	&	25	&	144.4	&	28.7	&	7.3	&	142	\\
\rm Collinder\,132\,gp\,2	&	107.042730	&	-25.625351	&&	689.6	&	-364.9	&	-576.7	&	-98.7	&	28.0  	&	-3.916	&	3.617	&	25	&	362.7	&	28.2	&	10.0	&	385	\\
\rm Collinder\,132\,gp\,3	&	107.437642	&	-25.401309	&&	602.2	&	-316.9	&	-506.1	&	-78.4	&	26.1  	&	-4.949	&	3.749	&	25	&	121.1	&	15.2	&	6.9	&	123	\\
\rm Collinder\,132\,gp\,4	&	107.181518	&	-30.328868	&&	789.1	&	-365.5	&	-686.4	&	-133.9	&	30.3  	&	-3.417	&	3.407	&	25	&	68.4	&	14.6	&	5.7	&	58	\\
\rm Collinder\,132\,gp\,5	&	107.638309	&	-27.792975	&&	599.4	&	-297.3	&	-513.2	&	-87.3	&	37.9  	&	-3.231	&	6.356	&	50	&	53.1	&	9.0	&	5.3	&	56	\\
\rm Collinder\,132\,gp\,6	&	111.643982	&	-30.322059	&&	615.6	&	-270.2	&	-548.8	&	-69.8	&	34.1  	&	-4.354	&	6.113	&	100	&	39.5	&	15.5	&	4.8	&	40	\\
\rm Gulliver\,21	&	106.972407	&	-25.450964	&&	648.7	&	-345.0	&	-542.1	&	-89.4	&	39.8  	&	-1.907	&	4.214	&	275	&	176.5	&	9.3	&	7.9	&	173	\\

\enddata
\tablecomments{
    Columns 2--10 list the median values of the subgroup member properties. R.A. and Decl. are the right ascension and declination. Dist. is the corrected distance. $X_c$, $Y_c$, and $Z_c$ are the positions of each subgroup in heliocentric Cartesian coordinates after distance correction. RV is the radial velocity. \pmra{} and \pmdec{} are the components of the proper motion. The age of each subgroup is derived from PARSEC isochrone fitting (Figure~\ref{fig:cmd}). $M_{\rm cl}$ is the total mass of each subgroup. $r_{\rm h}$ and $r_{\rm t}$ are the half-mass radius and the tidal radius of each subgroup, respectively. The tidal radius is computed using Equation~(12) in \citet{pinfield1998}. $N$ is the total number of members in each subgroup.
    }
\end{deluxetable*}

\section{Co-moving Populations}\label{sec:multi}
\subsection{Age and Metallicity Difference}\label{sec:clean}

\begin{figure*}[tb!]
\centering
\includegraphics[angle=0, width=1.\textwidth]{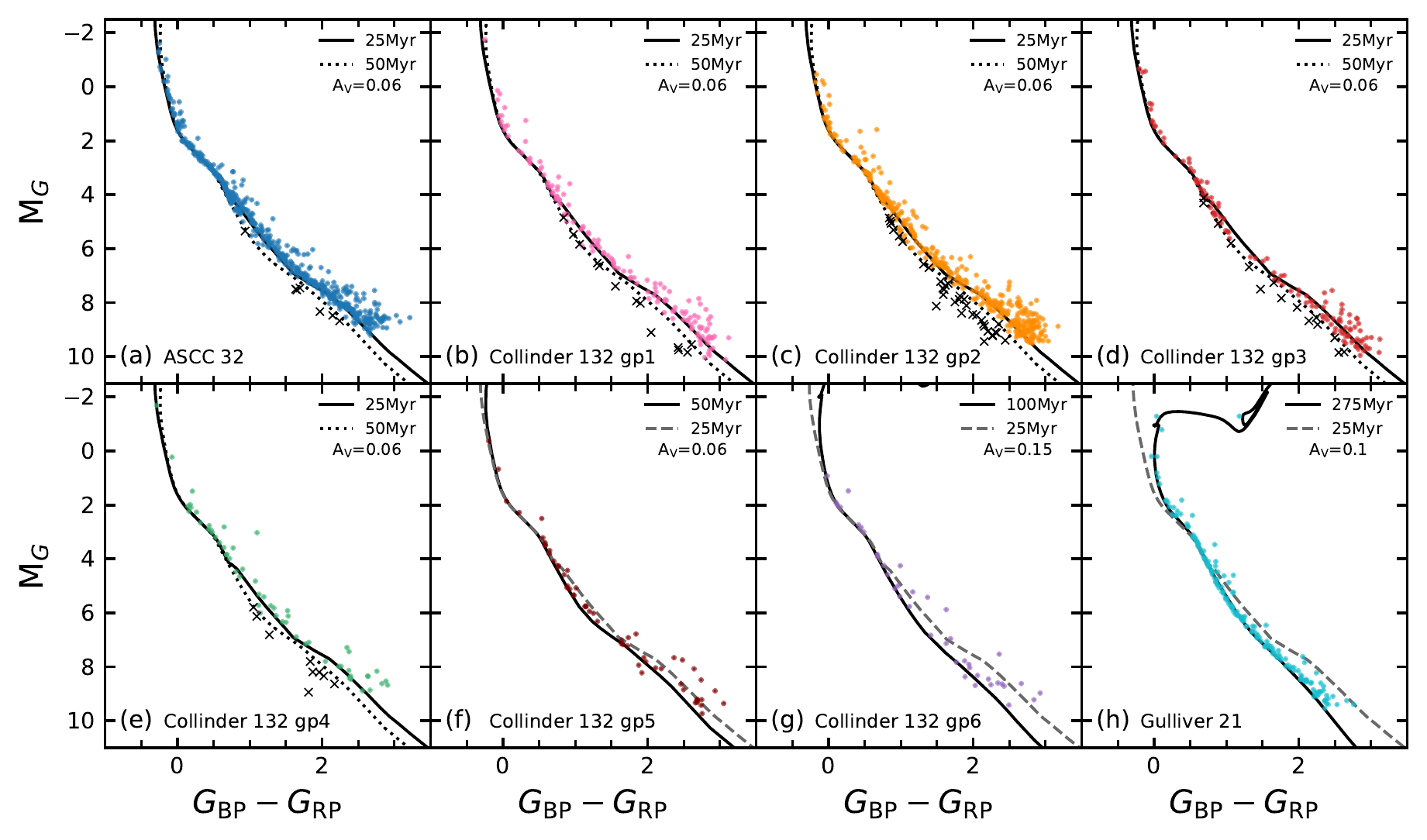}
    \caption{
    Absolute magnitude CMDs (with M$_G$ adopting Gaia DR\,3 parallaxes) for member stars obtained from Gaia DR\,3. The colored dots in each panel represent the corresponding member candidates in each subgroup. Black crossed symbols are field stars that are excluded from further investigation (see Section~\ref{sec:clean}). The PARSEC isochrones of the best fitted age are indicated with the black solid curves, with solar metallicity and estimated A$_{\rm V}$. The black dotted curve is isochrone of 50~Myr, with solar metallicity and A$_{\rm V}$=0.06. The grey dashed curves in (f)--(h) are 25\,Myr isochrones for comparison.
    }
\label{fig:cmd}
\end{figure*}

We construct color-magnitude diagrams (CMDs) for the member candidates of the eight subgroups of stars in the Collinder\,132-Gulliver\,21 regions. We first fit the PARSEC isochrone \citep[version 1.2S,][]{bressan2012,chen2015} to each subgroup by eye to estimate their age and reddening. All subgroups are assumed to have solar abundance. We observe three major populations of different ages among the eight subgroups in the target region (Figure~\ref{fig:cmd}). The first population is the Collinder\,132 moving group, which contains five subgroups with an age of 25\,Myr: ASCC\,32, Collinder\,132\,gp\,1--4. The second population is an intermediate-age groups (50--100\,Myr): Collinder\,132\,gp\,5--6. The third population is the oldest generation with an age of 275\,Myr: Gulliver\,21. The mass of each member star is then estimated from the best-fit isochrone using the k–D tree method \citep{mcmillan2007} by finding the nearest point on the isochrone.

In Figure~\ref{fig:cmd}~(a)--(e), a handful of candidate stars are located below the main sequence locus with age older than 50\,Myr (black dotted curve). Although these stars are located in the same sky region of the 25\,Myr-old Collinder\,132 moving group, we did not find any distinctive difference between these stars from the field stars nearby. Similar to the approach taken in Section~3.1 in \citet{pang2021b}, we consider stars bluer than the 50\,Myr-old isochrone and fainter than M$_G>4$\,mag as field stars contamination (black crossed symbols in panels (a)--(e) in Figure~\ref{fig:cmd}), and therefore exclude them from further analysis.

\begin{figure*}[tb!]
\centering
\includegraphics[angle=0, width=1.\textwidth]{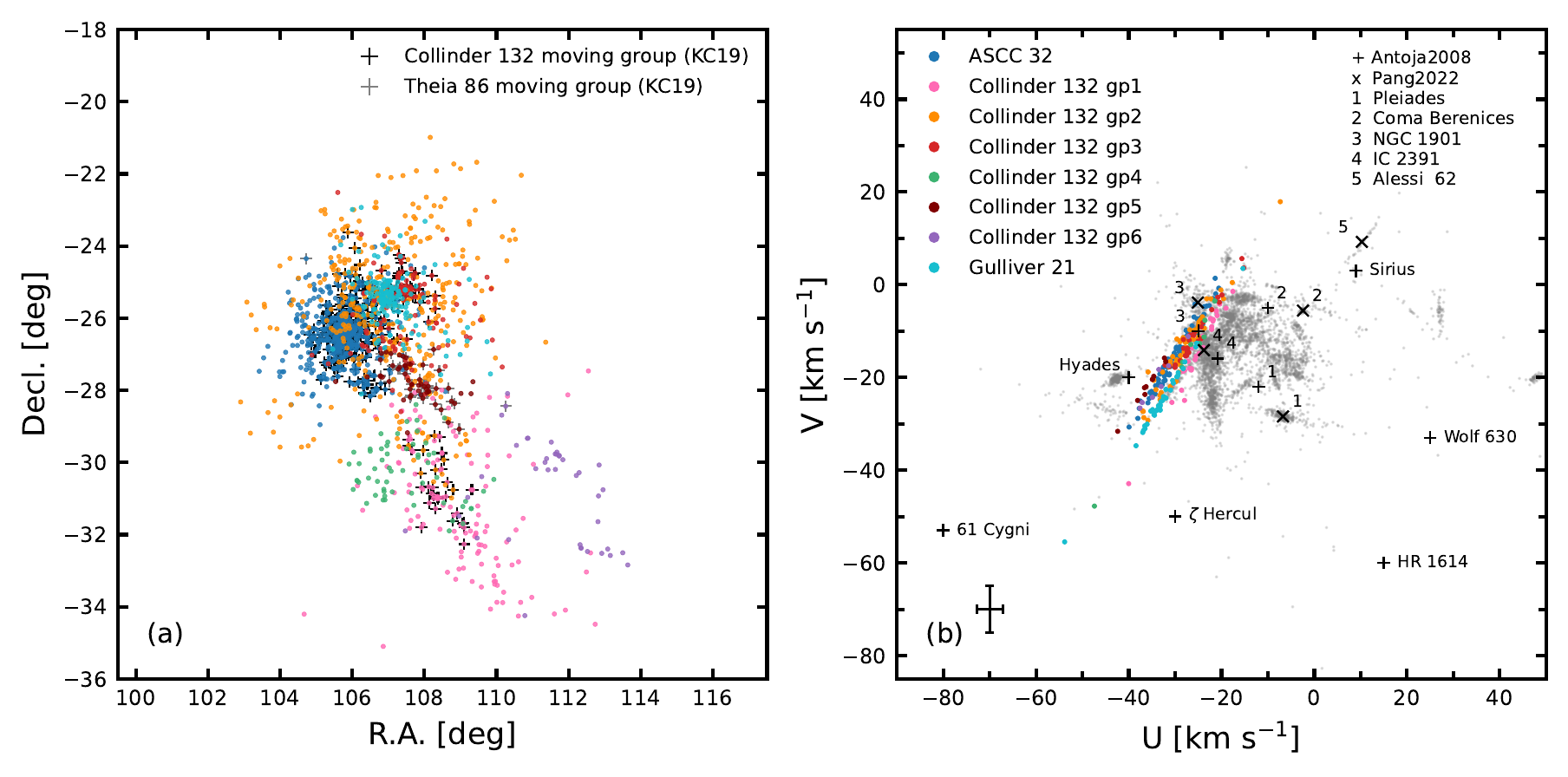}
    \caption{ 
    (a) Projected spatial distribution of member candidates. Cross-matched members with the Collinder\,132 moving group in \citet[][KC19]{kounkel2019} are indicated by the black plus symbols. Matched stars with Theia\,86 moving group in KC19 are shown as grey plus symbols.
    (b) The $U$--$V$ velocity distribution for eight subgroups in the Collinder\,132-Gulliver\,21 region obtained from this study. Members of these eight subgroups are shown as colored dots. The mean errors of $U$ and $V$ are indicated in the bottom-left corner.} The grey dots in the background are 85 clusters and groups from \citet{pang2022} for comparison.
    The S/N of RVs constructing velocities $U$ and $V$ are restricted to be greater than 20. The black cross indicates the median position of the open cluster Alessi\,62 taken from \citet{pang2022}. Alessi\,62 is a dissolving cluster with an age of 692\,Myr, which forms a diagonal stripe structure.
\label{fig:radec_uv}
\end{figure*}

A total of 1496 members of the eight stellar subgroups in the Collinder\,132-Gulliver\,21 region remain after CMD cleaning (Figure~\ref{fig:cmd}). Basic parameters of these eight subgroups are presented in Table~\ref{tab:general}. We provide a detailed member list of all subgroups in the Appendix Table~\ref{tab:memberlist}.  
Our list of members is in a good agreement with two moving groups studied in \citet{kounkel2019}: 332 stars in ASCC\,32 and Collinder\,132\,gp\,1--4 (black plus symbols in Figure~\ref{fig:radec_uv}~(a)) belong to the Collinder\,132 moving group in \citet{kounkel2019}; and 43 stars in Collinder\,132\,gp\,5--6 (grey plus symbols in Figure~\ref{fig:radec_uv} (a)) match with the members of the Theia\,86 moving group found by \citet{kounkel2019}. 
In our study we do not only double the number of member stars in the Collinder\,132 moving group compared to \citet{kounkel2019}, but we are also the first in identifying five hierarchical subgroups in the Collinder\,132 moving group.

The distribution of metallicities, [M/H], of 1351 identified members obtained from BP/RP spectra of Gaia DR\,3, indicates an abundance variation among the three populations. Five subgroups in the Collinder\,132 moving group have an almost homogeneous metallicity with a mean [M/H]$\sim-0.32$. The oldest population Gulliver\,21 has [M/H]$\sim-0.09$, different from the youngest generation by 0.3. The intermediate-age group has [M/H]$\sim-0.24$. Although the absolute value of metallicity from GSP-Phot has zero-point offset issues \citep{andrae_gaia_2022}, the relative values of [M/H] among three populations can still provide some hint in abundance discrepancy. Based on the 61 members with [M/H] measurements from the RVS in Gaia DR\,3, the [M/H] difference between Gulliver\,21 and the Collinder\,132 moving group ranges from 0.3 to 0.6 \citep[after calibration following][]{recio-blanco_gaia_2022}. The metallicities obtained from the BP/RP spectra are somewhat uncertain because of the low spectral resolution, and the ones obtained from the RVS are affected by low number statistics. Therefore, follow-up spectroscopy studies are needed to confirm the abundance discrepancy between Collinder\,132 and Gulliver\,21.

\subsection{Velocity Distribution Features}\label{sec:vel}

Identification of moving groups is normally carried out by searching for kinematic substructures in velocity space. RVs with a signal-to-noise ratio S/N$>$20 are selected to construct the velocity distribution to avoid bias by uncertainty. We compute the RV dispersion of each group using the Markov Chain Monte Carlo (MCMC) method \citep{pang2021a}, with a likelihood function for the RV distribution that is a combination of two Gaussian components: one for the cluster members and  one for the field stars \citep[equations~1 and~8 in][]{cottaar2012}. The resulting RV dispersion for these eight subgroups is in the range of $\sim$2.0--6.5\,\kmsnospace.

We present the three populations (eight subgroups) in the Collinder\,132-Gulliver\,21 region in the $U$--$V$ velocity plane (Heliocentric) in Figure~\ref{fig:radec_uv}(b). These three populations of stars all follow a tight correlation between $U$ and $V$, and form parallel diagonal stripes with identical slopes in the $U$--$V$ plane. The pattern of diagonal stripes  is robust since their full extension ($>30$\,\kmsnospace) is much larger than observational errors (indicated in the bottom-left corner of Figure~\ref{fig:radec_uv} (b)) and the RV dispersions.

The Coma Berenices, Hyades and Pleiades moving groups \citep[plus symbols in Figure~\ref{fig:radec_uv} (b),][]{antoja_origin_2008} surround the three populations of stars and the NGC\,1901 and IC\,2391 moving groups in a triangle pattern. Compared to the distribution of 85 open clusters and stellar groups from \citet[][grey dots in Figure~\ref{fig:radec_uv} (a)]{pang2022}, the kinematic structure of these three population
is similar to the 692\,Myr-old dissolving cluster Alessi\,62 (the crossed symbol No.5 in Figure~\ref{fig:radec_uv}~(b), in which stars are still comoving. Their $V$ values extend 30\,\kmsnospace, larger than any of the other clusters or groups from \citet{pang2022}, indicating their comoving state. In the $U$--$V$ plane, the moving groups (plus symbols) named after open clusters (Pleiades, Coma Berenices, NGC\,1901, and IC\,2391) appear to closely relate to the median velocities of the eponymous open clusters taken from \citet[][crossed symbols in Figure~\ref{fig:radec_uv} (b)]{pang2022}, with a small offset owning to different survey data and unequal member numbers between moving group and open clusters.  

The distinct substructures emerging from these three populations in the $U$--$V$ plane indicate that their kinematics resemble moving groups, which are assumed to be unbound. They are three individual co-moving populations: (i) a 25\,Myr-old Collinder\,132 moving group, (ii) an intermediate-age group (50--100\,Myr), and (iii) the oldest group Gulliver\,21 (275\,Myr).
Stars in each population are co-moving together. We call the kinematic structures generated by these three co-moving populations the {\em Collinder\,132-Gulliver\,21 stream}.

\section{Characteristics of Collinder\,132-Gulliver\,21 Stream}\label{sec:stream}
\subsection{Spatial Distribution}\label{sec:spa}

\begin{figure*}[tb!]
\centering
\includegraphics[angle=0, width=1.\textwidth]{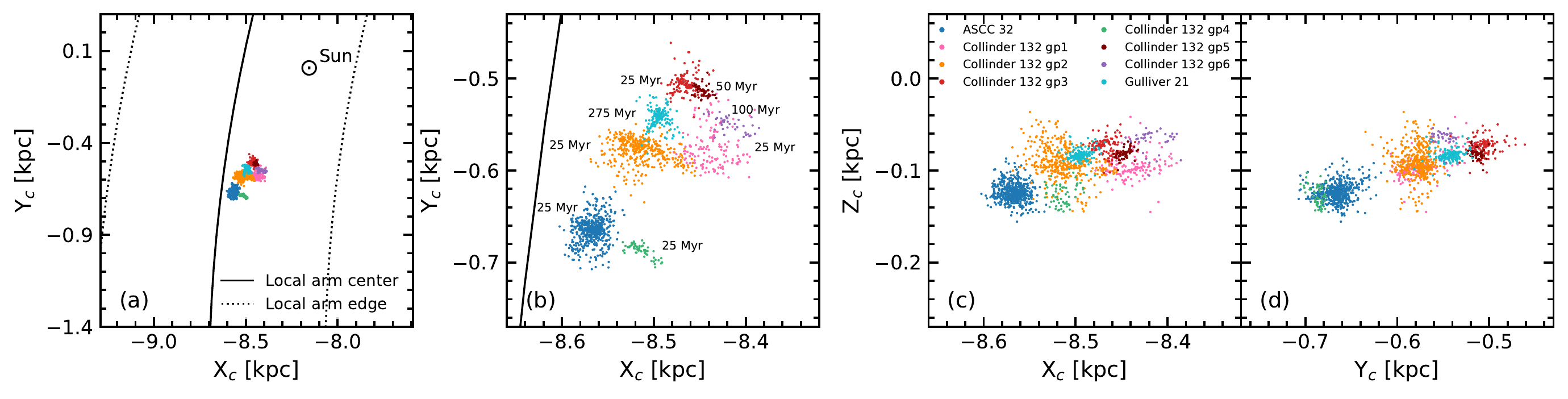}
    \caption{
    3D morphology of the Collinder\,132-Gulliver\,21 stream in Galactocentric Cartesian coordinates after distance correction via a Bayesian approach \citep{pang2021a}. The black solid curve represents the Local Arm center, and the Local Arm edge is denoted as the black dotted curve. The position of the Sun is taken at {\bf($X, Y, Z$) = ($-$8150, 0, 5.5)~pc} \citep{reid2019}. The age of each subgroup is indicated in panel (b) to show that the oldest population is surrounded by young populations.
    }
\label{fig:xyz}
\end{figure*}

The 3D spatial distribution shown in Figure~\ref{fig:xyz} (a), indicates that the Collinder\,132-Gulliver\,21 stream is located close to the Local Arm center \citep{reid2019}, where the stellar density is the highest. The entire Collinder\,132-Gulliver\,21 stream has a spatial extent of 270\,pc (Figure~\ref{fig:xyz} (b)--(d)) after distance correction, which followed the Bayesian method described in \citet{carrera2019} and \citet{pang2020,pang2021a}. The Collinder\,132 moving group and the intermediate-age groups encircle the oldest population Gulliver\,21 (Figure~\ref{fig:xyz} (b)). Subgroups in the Collinder\,132 moving group (ASCC\,32, Collinder\,132\,gp\,1--4) exhibit a filamentary morphology, similar to the filamentary-type stellar groups in \citet{pang2022}. This elongated filamentary shape of the Collinder\,132 moving group resembles the stellar relics of star formation \citep{jerabkova2019,beccari2020}. 
The abundance homogeneity (section~\ref{sec:clean}) in these five subgroups suggests that they all originate from a common GMC.

\begin{figure*}[tb!]
\centering
\includegraphics[angle=0, width=1.\textwidth]{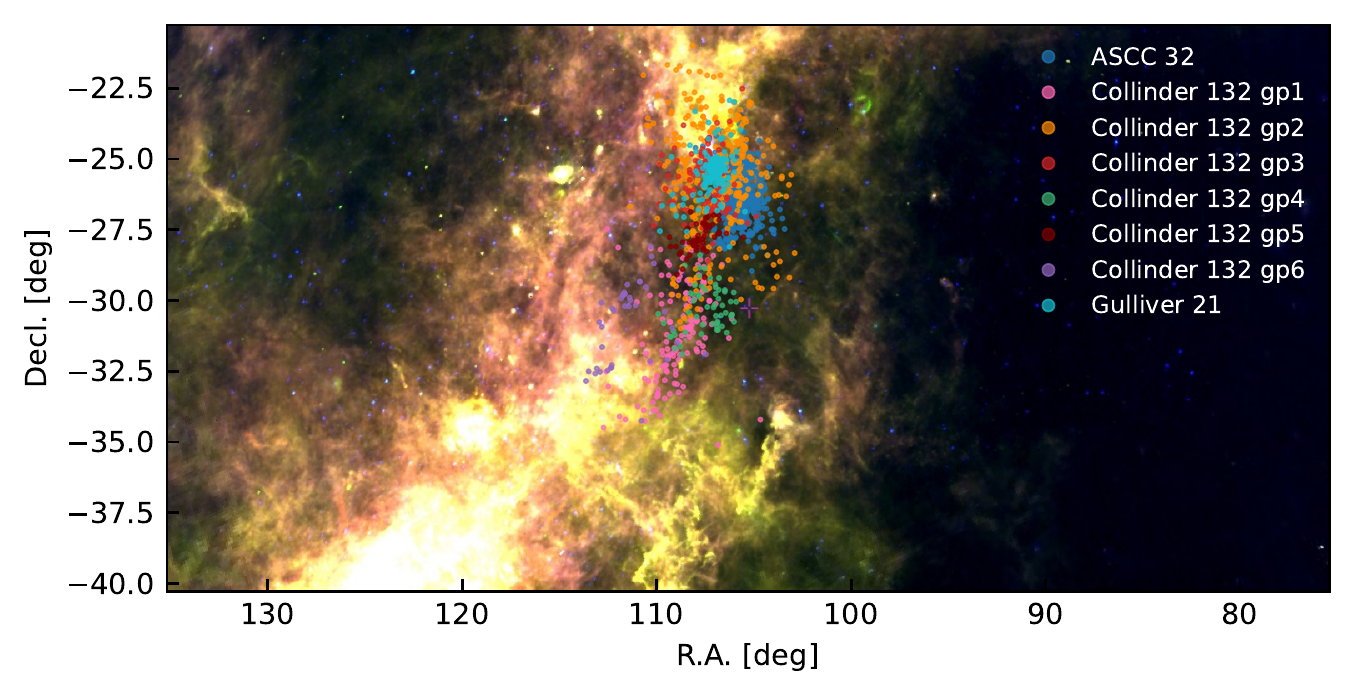}
    \caption{
    IRAS-IRIS infrared image of the 60~$\mu$m band \citep{miville2005}. Members of the eight subgroups in the Collinder\,132-Gulliver\,21 stream are displayed as colored dots. The background bubble structure is apparent.
    }
\label{fig:bubble_overplot}
\end{figure*}

To further investigate the star formation process in this region, we overlay the Collinder\,132-Gulliver\,21 stream on the IRIS \citep[Improved Reprocessing of the Survey image from the Infrared Astronomical Satellite, IRAS;][]{miville2005} image in Figure~\ref{fig:bubble_overplot}. The 60\,$\mu$m band image shows the gas structures in this region. A bubble-like structure is apparent in the background. This bubble has not been cataloged in existing bubble catalogs \citep{churchwell_bubbling_2006,churchwell2007,simpson_milky_2012,bania_arecibo_2012,hou2014,yan_molecular_2016} or the molecular cloud catalog \citep{chen2020} because it has a declination beyond the observation limit of these studies. \citet{dame2001} reported very weak CO emission from this region, indicating low gas density. The observed bubble in Figure~\ref{fig:bubble_overplot} should be located at a distance $>$800\,pc, based on the 3D extinction map, in which there is a sudden increase in the reddening at a distance beyond 800\,pc \citep{green2019}. Whether or not the bubble is adjacent to our target region is still unclear, as the members of ASCC\,32 extend beyond 800\,pc. The young Collinder\,132 moving group may have originated from a recent large-scale filamentary star formation episode, from which remaining gas may still be present. Gaia DR\,3 GSP-Phot extinctions in $G$, $G_{BP}$, $G_{RP}$ suggest an 0.1 mag higher reddening of the younger generations, compared to the oldest population, Gulliver\,21. Further investigation is needed to confirm this difference.

\subsection{Dynamical State}\label{sec:kinematics}


We adopt Equation~3 in \citet{fleck2006} with $\eta=9.75$ from \citet{pang2013} to estimate the dynamical mass of each subgroup in the Collinder\,132-Gulliver\,21 stream using the one-dimensional PM. These dynamical masses computed from Gaia DR\,3 proper motions for all subgroups are an order of magnitude larger than their corresponding photometric masses. In addition, the observed half-mass radii of these eight subgroups are all larger than their tidal radii (Table~\ref{tab:general}). Therefore, all three co-moving populations are gravitationally unbound and dissolving, living up to the name ``moving group''. The youngest moving group, Collinder\,132, probably experienced violent relaxation after a phase of significant gas expulsion \citep[e.g.,][]{baumgardt2007,pang2020}, as supported by the presence of the bubble in the region (Figure~\ref{fig:bubble_overplot}). 

\begin{figure*}[tb!]
\centering
\includegraphics[angle=0, width=1.\textwidth]{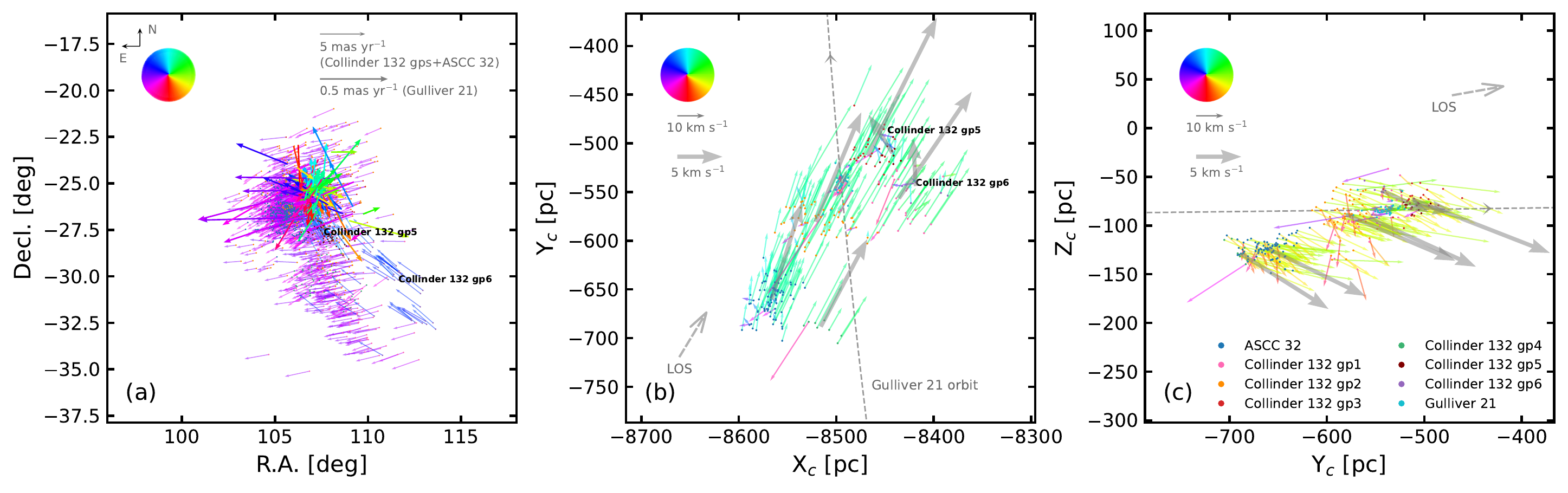}
    \caption{
    (a) The relative PMs of member candidates. All PM vectors are relative to the median PM of Gulliver\,21. The vectors of Gulliver\,21 members are thicker and are using different scaling than those for other groups (upper-right corner).
    (b)--(c) The relative 3D velocity vectors for members, projected onto the $X$--$Y$ and $Y$--$Z$ planes. The median motion of Gulliver\,21 is taken as the reference. We only show velocity vectors for stars with S/N$>$20 in RV, of which the mean error is 5.8\,\kmsnospace. The large thick grey arrows represent the mean velocity of each subgroup relative to that of Gulliver\,21, whose scaling is indicated in the upper-left corner (grey arrow)}. The grey dashed curve shows the orbit of Gulliver\,21, and the orbital direction is indicated with an arrow. The dashed arrows in (b) and (c) indicate the direction of the line of sight (LOS). The colors of the vectors in these three panels indicate the directions of the vectors, consistent with the color wheel. The scaling of the vectors is shown in each panel.
\label{fig:relative_velocity}
\end{figure*}

\section{Dynamical Origin of the Collinder\,132-Gulliver\,21 Stream}\label{sec:origin}

\subsection{Fly-by Scenario}\label{sec:fly}

To investigate the kinematical relation between each co-moving population in the Collinder\,132-Gulliver\,21 stream, we compute the relative PMs and 3D velocities for all subgroups with Gulliver\,21 median values as reference. In Figure~\ref{fig:relative_velocity}~(a), arrows are color-coded based on their pointing directions that match the color wheel on the upper left. Therefore, arrows with similar colors share similar relative velocities. Two distinct velocity populations can be identified: (i) a magenta arrow population pointing to the West, composed of mostly the young Collinder\,132 moving subgroups; and (ii) a blue arrow population pointing to the North-West, with intermediate-age subgroups. While the Collinder\,132 moving groups (magenta arrows) and the intermediate-age group (blue arrows) move in two unique directions with respect to Gulliver\,21, members of Gulliver\,21 show expansion patterns (thicker arrows). Similar signatures are observed in the distribution of relative 3D velocities (Figure~\ref{fig:relative_velocity} (b) and (c)). 
The mean 3D velocity of each subgroup, relative to Gulliver\,21, is represented with a large grey arrow (Figure~\ref{fig:relative_velocity} (b)), which is mostly larger than the RV dispersion ($\sim$2.0 to 6.5\,\kmsnospace)  and the mean RV error.
Gulliver\,21 is passing through the Collinder\,132 moving groups, while the intermediate-age group inclines to move toward Gulliver\,21's orbit.
Due to the weak gravitational forces between each subgroup, which are two orders of magnitude smaller than those of Galactic tidal forces, the trajectories of these three co-moving populations are not affected by their mutual gravitational interactions.

\begin{figure*}[tb!]
\centering
\includegraphics[angle=0, width=1.\textwidth]{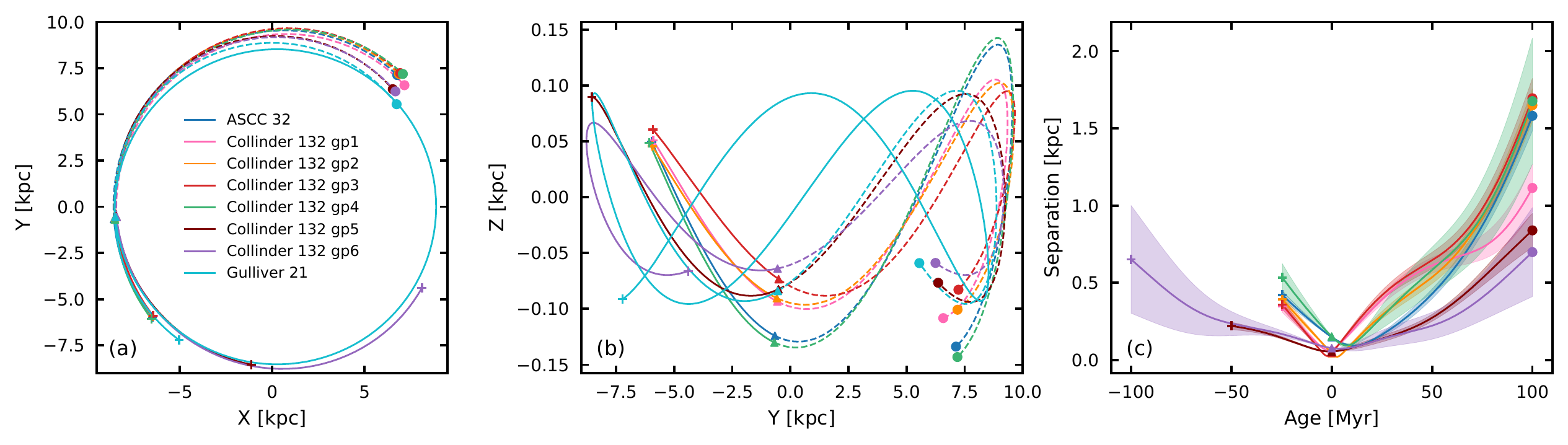}
    \caption{
    (a)--(b) Past and future 100~Myr integrated orbits of the eight subgroups in the Collinder\,132-Gulliver\,21 stream using \texttt{Galpy}. 
    (c) Evolution of the separation between seven subgroups and the oldest population, Gulliver\,21. The $\pm$1$\sigma$ uncertainty interval computed from observational errors in the PM and RV is indicated with the shaded areas. An age of 0~Myr corresponds to the present day. The colored plus symbols in the three panels indicate the positions of the corresponding subgroups at the time of birth. The triangles represent the present-day positions, and the filled circles indicate the predicated positions at 100~Myr from the present.
    }
\label{fig:galpy}
\end{figure*}

To disentangle the unique dynamics of each population, we integrate the orbits of eight subgroups back and forward in time, using the observed median value of the 3D velocities and 3D positions of each member in all subgroups. The publicly-available package \texttt{Galpy} \citep{Bovy2015} is used for the orbit integration. We adopt the value of 8.15\,kpc for the solar orbital distance, 247\,\kms for the Solar rotational velocity \citep{reid2019} and use the axis-symmetric Galactic potential model \texttt{MWPotential2014}, which is adequate for integrating the orbits of open clusters  \citep{wang2021a,pang2022,boffin2022}.  Figure~\ref{fig:galpy} (a) and (b) show the orbits of the eight subgroups in the three co-moving populations. The plus symbols indicate the positions of subgroups at their time of birth. The triangles are present-day positions, and the filled circles are the predicted positions at 100\,Myr in the future. All these eight subgroups follow almost circular orbits, with eccentricities ranging from 0.02 to 0.07. The distance between the seven younger subgroups and Gulliver\,21 reaches a minimum at the present time (Figure~\ref{fig:galpy} (c)) and will increase dramatically in the near future. 

The observed Collinder\,132-Gulliver\,21 stream is a transitional product of orbital overlap of the three co-moving populations. This temporary co-moving status can only exist for a period of about 70\,Myr, which is the estimated lifetime for the stream. This phase started approximately 20\,Myr ago. In approximately 50\,Myr from now, the Collinder\,132 moving groups, and the intermediate-age group will separate from Gulliver\,21. The Collinder\,132-Gulliver\,21 stream will therefore disappear. Even when observational errors are considered in calculating the distance between moving groups (shaded areas in Figure~\ref{fig:galpy} (c) show the $\pm$1 sigma interval), subgroups in each co-moving population will remain close to each other for at least 100\,Myr (Figure~\ref{fig:galpy} (c)). 

\subsection{Resonance Scenario}\label{sec:reso}

Known classical moving groups form a horizontal branch/arch of constant $V$ in the $U$--$V$ space, implying conservation of angular momentum along vertical direction, which is expected in the resonance scenario \citep{dehnen_effect_2000,bovy_velocity_2010,barros_exploring_2020}. However, the overdensities of the three co-moving populations in the Collinder\,132-Gulliver\,21 stream show ``parallelly tilted'' features on the $U$--$V$ velocity space. A similar feature has been observed in the moving group HR\,1614 \citep{kushniruk_hr_2020}, which is suggested to be induced by a combination mechanism of resonances and phase mixing. 

The Galactic bar resonance \citep{dehnen_effect_2000,bovy_velocity_2010}, the resonant scattering by transient spiral structure \citep{sellwood2002}, and the stochastic spiral wave \citep{s_de_simone_stellar_2004} are all able to change the stellar kinematics in the Galactic plane. As the Collinder\,132-Gulliver\,21 stream is located at the center of the Local Arm (Figure~\ref{fig:xyz} (a)), excitation and heating events from resonance could be frequent. Gulliver\,21 and the intermediate-age group might be affected by these mechanisms, and may have been scattered to the position of the Collinder\,132 moving group. Afterward, these three co-moving populations began to fly by each other and form the Collinder\,132-Gulliver\,21 stream about 20\,Myr ago. In approximately 50\,Myr from now, their orbits will diverge again.

\subsection{Phase-mixing Scenario}\label{sec:pha}

Phase mixing due to external perturbations  \citep{antoja_dynamically_2018,barros_exploring_2020,li2020,liz2021}, such as the previous pericentric passage of the Sagittarius dwarf galaxy across the Milky Way disk, can also produce similar kinematic structures like moving groups in the velocity space (i.e., $U$--$V$), and affects the vertical motion of stars and stars' position in the Galactic disk. Evidence of the vertical perturbation is reflected as the snail shell shape in the $Z-V_{Z}$ distribution for stars across the Galactic disk. However, previous studies have suggested that the perturbation occurred $\sim$500--700\,Myr ago \citep{li2020,liz2021,liw2021}, which is much earlier than the time at which the 25\,Myr-old Collinder\,132 moving groups were formed. Therefore, vertical phase mixing has had little effect on the stellar motions in the Collinder\,132-Gulliver\,21 stream. It is thus unlikely the cause for the formation of the Collinder\,132-Gulliver\,21 stream. On the other hand, the horizontal phase mixing of the stars in the Galactic disk could have been stimulated by the previous vertical perturbation, affecting the in-plane motion of the stars. The horizontal phase mixing might have triggered the mixture of Gulliver\,21, the intermediate group, and the Collinder\,132 moving group, resulting in the formation of the stream.

\section{Summary}\label{sec:summary}

In this study we have used Gaia DR\,3 data to study the Collinder\,132-Gulliver\,21 region, with the motivation to identify the relationship between the 275\,Myr-old Gulliver\,21 and the 25\,Myr-old Collinder\,132 moving group. Eight subgroups of stars are identified in the target region using \textsc{StarGO}. Our results can be summarized as follows:
\begin{enumerate}
    \item Eight subgroups in the region of Collinder\,132 and Gulliver\,21 are divided into three co-moving populations: (i) the youngest 25\,Myr-old Collinder\,132 moving group (ASCC\,32, Collinder\,132\,gp\,1--4); (ii) an intermediate-age group, consisting of Collinder\,132\,gp\,5--6 (50--100\,Myr); and (iii) the oldest population, Gulliver\,21 (275\,Myr). 

    \item In the $U$--$V$ velocity distribution, three populations stands out as parallel diagonal stripes, 
    following a tight $U$--$V$ correlation (with identical slope) with a $V$ difference of 30\,\kmsnospace. The elongated over-density generated by these three populations in the $U$--$V$ space is unique, compared to clusters or groups in the solar neighborhood. It resembles the kinematic structures of moving groups and indicates a co-moving state for each population. We name the kinematic structure formed by these three moving groups the {\em Collinder\,132-Gulliver\,21 stream}. 
    
    \item The 3D spatial distribution of the Collinder\,132-Gulliver\,21 stream extends 270\,pc. The oldest population, Gulliver\,21, is spatially surrounded by the Collinder\,132 moving group and the intermediate-age group. All three co-moving populations are gravitationally unbound, and are undergoing disruption.
    
    \item The Collinder\,132-Gulliver\,21 stream has a dynamical origin. The young  Collinder\,132 moving group was born in the spiral arm from filamentary star formation in its natal GMC. The Galactic bar and spiral structure resonance may then have scattered Gulliver\,21 and the intermediate-age group towards the location of the Collinder\,132 moving group. Three populations began to co-move as their orbits overlapped. After 50\,Myr from the present time, three co-moving populations in the Collinder\,132-Gulliver\,21 stream will start to separate. The stream will eventually disappear. Stars in each population will continue co-move for (at least) another 100\,Myr, and will subsequently become three separate moving groups.

\end{enumerate}

The orbit integration carried out in this work is based on axis-symmetric Galactic potential. A time-varying potential with perturbation is worth investigating in the future. Although there are indications of abundance variation of 0.3~dex between the youngest Collinder\,132 moving group and the oldest population Gulliver\,21, high-resolution spectroscopy is required to verify the inhomogeneous metallicity in the Collinder\,132-Gulliver\,21 stream.

\clearpage{}
\appendix{}
\counterwithin{figure}{section}
\counterwithin{table}{section}

\section{Member Selection with \textsc{StarGO}} \label{sec:som}

The \textsc{StarGO} \citep[STARs’ Galactic Origin,][]{yuan2018} software, based on the Self-Organizing Map (SOM) algorithm, is able to map high-dimensional data onto a 2D neural network and to search for groupings in the multi-dimensional parameters space.
The SOM algorithm works as follows; first, a 2D neural network is generated with either 100$\times$100 or 150$\times$150 neurons, depending on the number of input stars. Second, each neuron is given a random vector that matches the dimension of the input data, i.e., a 5D weight vector that matches with the 5D input data set, $X$, $Y$, $Z$, \pmra, and \pmdec, is generated. Third, we train a 2D neural network by feeding the input data (stars) one by one to the 2D neural network. \textsc{StarGO} will identify the neuron whose 5D weight vector is closest to the input data. This identified neuron and its neighboring neurons will have their 5D weight vector updated to be closer to the associated input data. Each training iteration of the 2D neural network is finished after looping every star in the target sky region. After each training iteration, the 2D neural network will have patches of neurons sharing similar 5D weight vectors. This training cycle is set to iterate 400 times for convergence.

\begin{figure*}[b!]
\centering
\includegraphics[angle=0, width=0.8\textwidth]{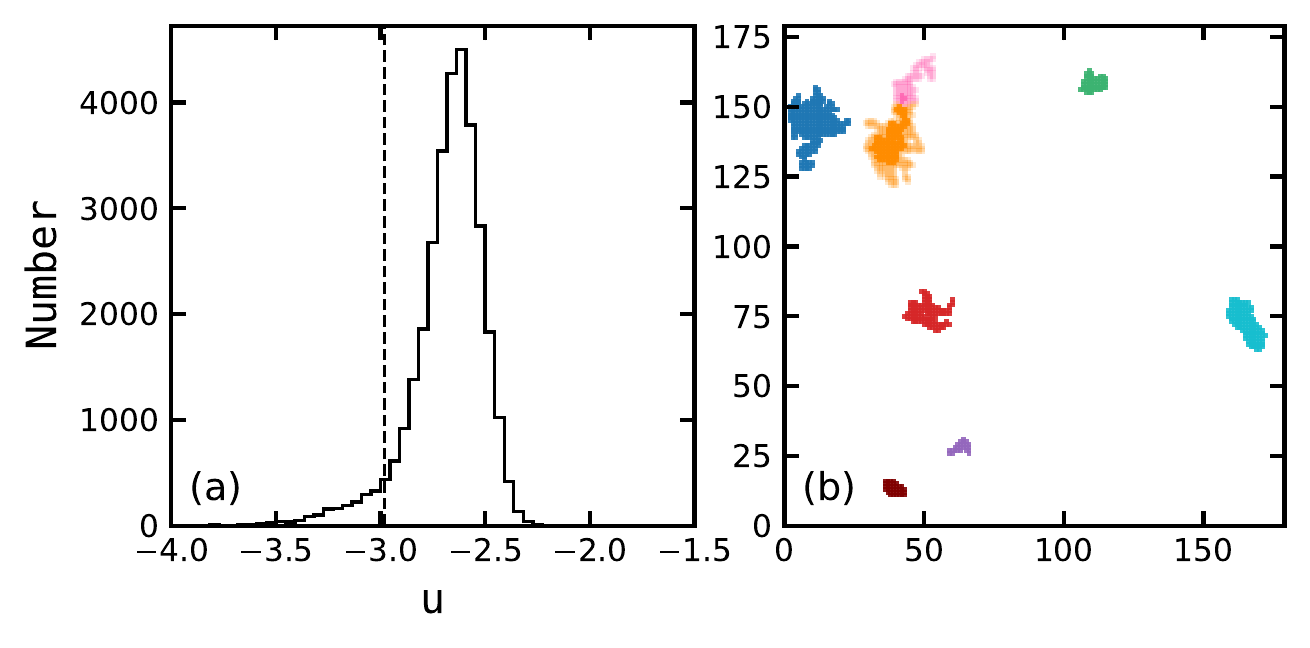}
    \caption{
    (a) Histogram of the distribution of $u$-matrix elements. The vertical dashed line denotes the threshold value of u that gives a 10\% contamination rate for the identified groups, corresponding to the color patches in the 2D neural network (panel (b)). (b) Neural network generated by SOM. Neurons with an $u$-threshold corresponding to a 10\% contamination rate are shown as colored patches. The transparent pink and orange patches are considered as the top-level structure overall. Solid pink and orange patches are the cores of the secondary hierarchical structures, which correspond to Collinder\,132\,gp\,1 (pink) and group 2 (orange). Other red, green, maroon, purple, and cyan patches correspond to Collinder\,132\,gp\,3, Collinder\,132\,gp\,4, Collinder\,132\,gp\,5, Collinder\,132\,gp\,6, and Gulliver\,21 with a contamination rate of 10\%, respectively.}
\label{fig:som}
\end{figure*}

The difference between the weight vectors of adjacent neurons on the 2D neural network is defended as the $u$-value. Therefore, the smaller the $u$-values, the higher the likelihood of stars associated to the neuron to be located in the same stellar group. Each input data (stars) is associated with the neuron which has a minimum difference in the 5D weight vector. Figure~\ref{fig:som} panel (a) shows the histogram of the $u$-value from the 2D neural network of the Collinder\,132-Gulliver\,21 sky region. The vertical dashed line shows the threshold cut, which gives a 10\% contamination rate for the identified groups. The field star contamination rate is estimated using the mock Gaia EDR\,3 catalog \citep{rybizki2020}. We process the field stars in the mock Gaia EDR\,3 catalog via the same procedure mentioned in the previous paragraph. Those mock stars attached to the patches of members in the 2D neural network trained by the observational data are considered as false positive, i.e., field star contaminants. Figure~\ref{fig:som} panel (b) show patches of neurons corresponding to a 10\% contamination rate.

\clearpage{}

\section{Member List} \label{sec:mem_list}

Table~\ref{tab:memberlist} provide a detailed member list of the Collinder\,132-Gulliver\,21 stream (Table~\ref{tab:memberlist}) determined by this study. 

\begin{deluxetable*}{ccc}
\tablecaption{Columns for the table of individual members of the eight subgroups in Collinder\,132-Gulliver\,21 stream. \label{tab:memberlist}
             }
\tabletypesize{\scriptsize}
\tablehead{
\colhead{Column}    & \colhead{Unit}    & \colhead{Description}
}
\startdata
Cluster Name                    & -                 &  Name of the target cluster   \\
      Gaia ID                   & -                 &  Object ID in Gaia DR\,3\\
ra                              & degree           &  R.A. at J2016.0 from Gaia DR\,3\\
er\_RA                          & mas              &  Positional uncertainty in R.A. at J2016.0 \\
dec                             & degree           &  Decl. at J2016.0 from Gaia DR\,3 \\
er\_DEC                         & mas              &  Positional uncertainty in decl. at J2016.0 \\
parallax                        & mas              &  Parallax from Gaia DR\,3\\
er\_parallax                    & mas              &  Uncertainty in the parallax \\
pmra                            & mas~yr$^{-1}$    &  Proper motion with robust fit in $\alpha \cos\delta$ from {\it Gaia} DR\,3     \\
er\_pmra                        & mas~yr$^{-1}$    &  Error of the proper motion with robust fit in $\alpha \cos\delta$   \\
pmdec                           & mas~yr$^{-1}$    &  Proper motion with robust fit in $\delta$ from Gaia DR\,3     \\
er\_pmdec                       & mas~yr$^{-1}$    &  Error of the proper motion with robust fit in $\delta$  \\
Gmag                            & mag              & Magnitude in $G$ band from Gaia DR\,3   \\
BP                              & mag              & Magnitude in $BP$ band from Gaia DR\,3   \\
RP                              & mag              & Magnitude in $RP$ band from Gaia DR\,3   \\
Gaia\_radial\_velocity          & km~s$^{-1}$      &  Radial velocity from Gaia DR\,3 \\
er\_Gaia\_radial\_velocity      & km~s$^{-1}$      &  Error of radial velocity from Gaia DR\,3\\
Mass                            & M$_\odot$        & Stellar mass obtained in this study\\
X\_obs                          & pc               & Heliocentric Cartesian X coordinate computed via direct inverting Gaia DR\,3 parallax $\varpi$ \\
Y\_obs                          & pc               & Heliocentric Cartesian Y coordinate computed via direct inverting Gaia DR\,3 parallax $\varpi$ \\
Z\_obs                          & pc               & Heliocentric Cartesian Z coordinate computed via direct inverting Gaia DR\,3 parallax $\varpi$ \\
X\_cor                          & pc               & Heliocentric Cartesian X coordinate after distance correction in this study \\
Y\_cor                          & pc               & Heliocentric Cartesian Y coordinate after distance correction in this study \\
Z\_cor                          & pc               & Heliocentric Cartesian Z coordinate after distance correction in this study \\
Dist\_cor                       & pc               & The corrected distance of individual member\\
\enddata
\tablecomments{A machine-readable version of this table is available online.
    }
\end{deluxetable*}

\clearpage
\begin{acknowledgments}
We thank the anonymous referee for their constructive comments and suggestions that helped to improve this paper.
This work is supported by the grant of National Natural Science Foundation of China, No: 12173029. 
Xiaoying Pang acknowledges the financial support of the research development fund of Xi'an Jiaotong-Liverpool University (RDF-18--02--32). 
L.W. thanks the support from the one-hundred-talent project of Sun Yat-sen University, the Fundamental Research Funds for the Central Universities, Sun Yat-sen University (22hytd09) and the National Natural Science Foundation of China through grant 12073090. 
M.B.N.K. acknowledges support from the National Natural Science Foundation of China (grant 11573004) and Xi'an Jiaotong-Liverpool University (grant RDF-SP-93). 
Z.Y.L. is supported by the National Natural Science Foundation of China under grant No. 12122301, by a Shanghai Natural Science Research Grant (21ZR1430600), by the Cultivation Project for LAMOST Scientific Payoff and Research Achievement of CAMS-CAS, by the ``111'' project of the Ministry of Education under grant No. B20019.

This work made use of data from the European Space Agency (ESA) mission {\it Gaia} 
(\url{https://www.cosmos.esa.int/gaia}), processed by the {\it Gaia} Data Processing 
and Analysis Consortium (DPAC, \url{https://www.cosmos.esa.int/web/gaia/dpac/consortium}). This study also made use of 
the SIMBAD database and the VizieR catalogue access tool, both operated at CDS, Strasbourg, France.
\end{acknowledgments}


\software{  \texttt{Astropy} \citep{astropy2013,astropy2018}, 
            \texttt{SciPy} \citep{millman2011},
            \texttt{TOPCAT} \citep{taylor2005}, and 
            \textsc{StarGO} \citep{yuan2018}.
}
\bibliography{main}

\begin{thebibliography}{}
\expandafter\ifx\csname natexlab\endcsname\relax\def\natexlab#1{#1}\fi
\providecommand{\url}[1]{\href{#1}{#1}}
\providecommand{\dodoi}[1]{doi:~\href{http://doi.org/#1}{\nolinkurl{#1}}}
\providecommand{\doeprint}[1]{\href{http://ascl.net/#1}{\nolinkurl{http://ascl.net/#1}}}
\providecommand{\doarXiv}[1]{\href{https://arxiv.org/abs/#1}{\nolinkurl{https://arxiv.org/abs/#1}}}

\bibitem[{Andrae {et~al.}(2022)Andrae, Fouesneau, Sordo, Bailer-Jones,
  Dharmawardena, Rybizki, De~Angeli, Lindstrom, Marshall, Drimmel, Korn,
  Soubiran, Brouillet, Casamiquela, \& {CU8 team}}]{andrae_gaia_2022}
Andrae, R., Fouesneau, M., Sordo, R., {et~al.} 2022, Astronomy \& Astrophysics,
  \dodoi{10.1051/0004-6361/202243462}

\bibitem[{Antoja {et~al.}(2008)Antoja, Figueras, Fernández, \&
  Torra}]{antoja_origin_2008}
Antoja, T., Figueras, F., Fernández, D., \& Torra, J. 2008, Astronomy \&
  Astrophysics, 490, 135, \dodoi{10.1051/0004-6361:200809519}

\bibitem[{Antoja {et~al.}(2012)Antoja, Helmi, Bienayme, Bland-Hawthorn, Famaey,
  Freeman, Gibson, Gilmore, Grebel, Minchev, Munari, Navarro, Parker, Reid,
  Seabroke, Siebert, Siviero, Steinmetz, Williams, Wyse, \&
  Zwitter}]{antoja_kinematic_2012}
Antoja, T., Helmi, A., Bienayme, O., {et~al.} 2012, Monthly Notices of the
  Royal Astronomical Society: Letters, 426, L1,
  \dodoi{10.1111/j.1745-3933.2012.01310.x}

\bibitem[{Antoja {et~al.}(2018)Antoja, Helmi, Romero-Gomez, Katz, Babusiaux,
  Drimmel, Evans, Figueras, Poggio, Reyle, Robin, Seabroke, \&
  Soubiran}]{antoja_dynamically_2018}
Antoja, T., Helmi, A., Romero-Gomez, M., {et~al.} 2018, Nature, 561, 360,
  \dodoi{10.1038/s41586-018-0510-7}

\bibitem[{{Astropy Collaboration} {et~al.}(2013){Astropy Collaboration},
  {Robitaille}, {Tollerud}, {Greenfield}, {Droettboom}, {Bray}, {Aldcroft},
  {Davis}, {Ginsburg}, {Price-Whelan}, {Kerzendorf}, {Conley}, {Crighton},
  {Barbary}, {Muna}, {Ferguson}, {Grollier}, {Parikh}, {Nair}, {Unther},
  {Deil}, {Woillez}, {Conseil}, {Kramer}, {Turner}, {Singer}, {Fox}, {Weaver},
  {Zabalza}, {Edwards}, {Azalee Bostroem}, {Burke}, {Casey}, {Crawford},
  {Dencheva}, {Ely}, {Jenness}, {Labrie}, {Lim}, {Pierfederici}, {Pontzen},
  {Ptak}, {Refsdal}, {Servillat}, \& {Streicher}}]{astropy2013}
{Astropy Collaboration}, {Robitaille}, T.~P., {Tollerud}, E.~J., {et~al.} 2013,
  \aap, 558, A33, \dodoi{10.1051/0004-6361/201322068}

\bibitem[{{Astropy Collaboration} {et~al.}(2018){Astropy Collaboration},
  {Price-Whelan}, {Sip{\H{o}}cz}, {G{\"u}nther}, {Lim}, {Crawford}, {Conseil},
  {Shupe}, {Craig}, {Dencheva}, {Ginsburg}, {VanderPlas}, {Bradley},
  {P{\'e}rez-Su{\'a}rez}, {de Val-Borro}, {Aldcroft}, {Cruz}, {Robitaille},
  {Tollerud}, {Ardelean}, {Babej}, {Bach}, {Bachetti}, {Bakanov}, {Bamford},
  {Barentsen}, {Barmby}, {Baumbach}, {Berry}, {Biscani}, {Boquien}, {Bostroem},
  {Bouma}, {Brammer}, {Bray}, {Breytenbach}, {Buddelmeijer}, {Burke},
  {Calderone}, {Cano Rodr{\'\i}guez}, {Cara}, {Cardoso}, {Cheedella}, {Copin},
  {Corrales}, {Crichton}, {D'Avella}, {Deil}, {Depagne}, {Dietrich}, {Donath},
  {Droettboom}, {Earl}, {Erben}, {Fabbro}, {Ferreira}, {Finethy}, {Fox},
  {Garrison}, {Gibbons}, {Goldstein}, {Gommers}, {Greco}, {Greenfield},
  {Groener}, {Grollier}, {Hagen}, {Hirst}, {Homeier}, {Horton}, {Hosseinzadeh},
  {Hu}, {Hunkeler}, {Ivezi{\'c}}, {Jain}, {Jenness}, {Kanarek}, {Kendrew},
  {Kern}, {Kerzendorf}, {Khvalko}, {King}, {Kirkby}, {Kulkarni}, {Kumar},
  {Lee}, {Lenz}, {Littlefair}, {Ma}, {Macleod}, {Mastropietro}, {McCully},
  {Montagnac}, {Morris}, {Mueller}, {Mumford}, {Muna}, {Murphy}, {Nelson},
  {Nguyen}, {Ninan}, {N{\"o}the}, {Ogaz}, {Oh}, {Parejko}, {Parley}, {Pascual},
  {Patil}, {Patil}, {Plunkett}, {Prochaska}, {Rastogi}, {Reddy Janga},
  {Sabater}, {Sakurikar}, {Seifert}, {Sherbert}, {Sherwood-Taylor}, {Shih},
  {Sick}, {Silbiger}, {Singanamalla}, {Singer}, {Sladen}, {Sooley},
  {Sornarajah}, {Streicher}, {Teuben}, {Thomas}, {Tremblay}, {Turner},
  {Terr{\'o}n}, {van Kerkwijk}, {de la Vega}, {Watkins}, {Weaver}, {Whitmore},
  {Woillez}, {Zabalza}, \& {Astropy Contributors}}]{astropy2018}
{Astropy Collaboration}, {Price-Whelan}, A.~M., {Sip{\H{o}}cz}, B.~M., {et~al.}
  2018, \aj, 156, 123, \dodoi{10.3847/1538-3881/aabc4f}

\bibitem[{Bania {et~al.}(2012)Bania, Anderson, \& Balser}]{bania_arecibo_2012}
Bania, T.~M., Anderson, L.~D., \& Balser, D.~S. 2012, The Astrophysical
  Journal, 14

\bibitem[{Barros {et~al.}(2020)Barros, Pérez-Villegas, Lépine, Michtchenko,
  \& Vieira}]{barros_exploring_2020}
Barros, D.~A., Pérez-Villegas, A., Lépine, J. R.~D., Michtchenko, T.~A., \&
  Vieira, R. S.~S. 2020, The Astrophysical Journal, 888, 75,
  \dodoi{10.3847/1538-4357/ab59d1}

\bibitem[{{Baumgardt} \& {Kroupa}(2007)}]{baumgardt2007}
{Baumgardt}, H., \& {Kroupa}, P. 2007, \mnras, 380, 1589,
  \dodoi{10.1111/j.1365-2966.2007.12209.x}

\bibitem[{{Beccari} {et~al.}(2020){Beccari}, {Boffin}, \&
  {Jerabkova}}]{beccari2020}
{Beccari}, G., {Boffin}, H. M.~J., \& {Jerabkova}, T. 2020, \mnras, 491, 2205,
  \dodoi{10.1093/mnras/stz3195}

\bibitem[{Bensby {et~al.}(2014)Bensby, Feltzing, \&
  Oey}]{bensby_exploring_2014}
Bensby, T., Feltzing, S., \& Oey, M.~S. 2014, Astronomy \& Astrophysics, 562,
  A71, \dodoi{10.1051/0004-6361/201322631}

\bibitem[{Bensby {et~al.}(2007)Bensby, Oey, Feltzing, \&
  Gustafsson}]{bensby_disentangling_2007}
Bensby, T., Oey, M.~S., Feltzing, S., \& Gustafsson, B. 2007, The Astrophysical
  Journal, 655, L89, \dodoi{10.1086/512014}

\bibitem[{{Boffin} {et~al.}(2022){Boffin}, {Jerabkova}, {Beccari}, \&
  {Wang}}]{boffin2022}
{Boffin}, H. M.~J., {Jerabkova}, T., {Beccari}, G., \& {Wang}, L. 2022, \mnras,
  514, 3579, \dodoi{10.1093/mnras/stac1567}

\bibitem[{{Bovy}(2015)}]{Bovy2015}
{Bovy}, J. 2015, \apjs, 216, 29, \dodoi{10.1088/0067-0049/216/2/29}

\bibitem[{Bovy \& Hogg(2010)}]{bovy_velocity_2010}
Bovy, J., \& Hogg, D.~W. 2010, The Astrophysical Journal, 717, 617,
  \dodoi{10.1088/0004-637X/717/2/617}

\bibitem[{{Bressan} {et~al.}(2012){Bressan}, {Marigo}, {Girardi}, {Salasnich},
  {Dal Cero}, {Rubele}, \& {Nanni}}]{bressan2012}
{Bressan}, A., {Marigo}, P., {Girardi}, L., {et~al.} 2012, \mnras, 427, 127,
  \dodoi{10.1111/j.1365-2966.2012.21948.x}

\bibitem[{{Cantat-Gaudin} {et~al.}(2018){Cantat-Gaudin}, {Jordi}, {Vallenari},
  {Bragaglia}, {Balaguer-N{\'u}{\~n}ez}, {Soubiran}, {Bossini}, {Moitinho},
  {Castro-Ginard}, {Krone-Martins}, {Casamiquela}, {Sordo}, \&
  {Carrera}}]{cantat2018}
{Cantat-Gaudin}, T., {Jordi}, C., {Vallenari}, A., {et~al.} 2018, \aap, 618,
  A93, \dodoi{10.1051/0004-6361/201833476}

\bibitem[{{Cantat-Gaudin} {et~al.}(2020){Cantat-Gaudin}, {Anders},
  {Castro-Ginard}, {Jordi}, {Romero-Gomez}, {Soubiran}, {Casamiquela},
  {Tarricq}, {Moitinho}, {Vallenari}, {Bragaglia}, {Krone-Martins}, \&
  {Kounkel}}]{Cantat-Gaudin2020}
{Cantat-Gaudin}, T., {Anders}, F., {Castro-Ginard}, A., {et~al.} 2020, VizieR
  Online Data Catalog, J/A+A/640/A1

\bibitem[{{Carrera} {et~al.}(2019){Carrera}, {Pasquato}, {Vallenari},
  {Balaguer-N{\'u}{\~n}ez}, {Cantat-Gaudin}, {Mapelli}, {Bragaglia}, {Bossini},
  {Jordi}, {Galad{\'\i}-Enr{\'\i}quez}, \& {Solano}}]{carrera2019}
{Carrera}, R., {Pasquato}, M., {Vallenari}, A., {et~al.} 2019, \aap, 627, A119,
  \dodoi{10.1051/0004-6361/201935599}

\bibitem[{{Chen} {et~al.}(2020){Chen}, {Li}, {Yuan}, {Huang}, {Tian}, {Wang},
  {Zhang}, {Wang}, \& {Liu}}]{chen2020}
{Chen}, B.~Q., {Li}, G.~X., {Yuan}, H.~B., {et~al.} 2020, \mnras, 493, 351,
  \dodoi{10.1093/mnras/staa235}

\bibitem[{{Chen} {et~al.}(2015){Chen}, {Bressan}, {Girardi}, {Marigo}, {Kong},
  \& {Lanza}}]{chen2015}
{Chen}, Y., {Bressan}, A., {Girardi}, L., {et~al.} 2015, \mnras, 452, 1068,
  \dodoi{10.1093/mnras/stv1281}

\bibitem[{Churchwell {et~al.}(2006)Churchwell, Povich, Allen, Taylor, Meade,
  Babler, Indebetouw, Watson, Whitney, Wolfire, Bania, Benjamin, Clemens,
  Cohen, Cyganowski, Jackson, Kobulnicky, Mathis, Mercer, Stolovy, Uzpen,
  Watson, \& Wolff}]{churchwell_bubbling_2006}
Churchwell, E., Povich, M.~S., Allen, D., {et~al.} 2006, The Astrophysical
  Journal, 649, 759, \dodoi{10.1086/507015}

\bibitem[{{Churchwell} {et~al.}(2007){Churchwell}, {Watson}, {Povich},
  {Taylor}, {Babler}, {Meade}, {Benjamin}, {Indebetouw}, \&
  {Whitney}}]{churchwell2007}
{Churchwell}, E., {Watson}, D.~F., {Povich}, M.~S., {et~al.} 2007, \apj, 670,
  428, \dodoi{10.1086/521646}

\bibitem[{{Clari{\'a}}(1977)}]{claria1977}
{Clari{\'a}}, J.~J. 1977, \pasp, 89, 803, \dodoi{10.1086/130231}

\bibitem[{{Collinder}(1931)}]{collinder1931}
{Collinder}, P. 1931, Annals of the Observatory of Lund, 2, B1

\bibitem[{{Cottaar} {et~al.}(2012){Cottaar}, {Meyer}, \&
  {Parker}}]{cottaar2012}
{Cottaar}, M., {Meyer}, M.~R., \& {Parker}, R.~J. 2012, \aap, 547, A35,
  \dodoi{10.1051/0004-6361/201219673}

\bibitem[{Creevey {et~al.}(2022)Creevey, Sordo, Pailler, Frémat, Heiter,
  Thévenin, Andrae, Fouesneau, Lobel, Bailer-Jones, Garabato, Bellas-Velidis,
  Brugaletta, Lorca, Ordenovic, Palicio, Sarro, Delchambre, Drimmel, Rybizki,
  Elipe, Korn, Recio-Blanco, Schultheis, De~Angeli, Montegriffo, Aramburu,
  Accart, Álvarez, Bakker, Brouillet, Burlacu, Carballo, Casamiquela,
  Chiavassa, Contursi, Cooper, Dafonte, Dapergolas, de~Laverny, Dharmawardena,
  Edvardsson, Fustec, García-Lario, García-Torres, Gomez,
  González-Santamaría, Hatzidimitriou, Piccolo, Kontizas, Kordopatis,
  Lanzafame, Lebreton, Licata, Lindstrøm, Livanou, Romeo, Manteiga, Marocco,
  Marshall, Mary, Nicolas, Pallas-Quintela, Panem, Pichon, Poggio, Riclet,
  Robin, Santoveña, Silvelo, Slezak, Smart, Soubiran, Süveges, Ulla, Utrilla,
  Vallenari, Zhao, Zorec, Barrado, Bijaoui, Bouret, Blomme, Brott, Cassisi,
  Kochukhov, Martayan, Shulyak, \& Silvester}]{creevey_gaia_2022}
Creevey, O.~L., Sordo, R., Pailler, F., {et~al.} 2022, Gaia {Data} {Release} 3:
  {Astrophysical} parameters inference system ({Apsis}) {I} -- methods and
  content overview,  arXiv.
\newblock \url{http://arxiv.org/abs/2206.05864}

\bibitem[{{Dame} {et~al.}(2001){Dame}, {Hartmann}, \& {Thaddeus}}]{dame2001}
{Dame}, T.~M., {Hartmann}, D., \& {Thaddeus}, P. 2001, \apj, 547, 792,
  \dodoi{10.1086/318388}

\bibitem[{Dehnen(2000)}]{dehnen_effect_2000}
Dehnen, W. 2000, The Astronomical Journal, 119, 800, \dodoi{10.1086/301226}

\bibitem[{{Eggen}(1983)}]{eggen1983}
{Eggen}, O.~J. 1983, \aj, 88, 197, \dodoi{10.1086/113306}

\bibitem[{{Eggen}(1996)}]{eggen1996}
---. 1996, \aj, 112, 1595, \dodoi{10.1086/118126}

\bibitem[{{Fleck} {et~al.}(2006){Fleck}, {Boily}, {Lan{\c{c}}on}, \&
  {Deiters}}]{fleck2006}
{Fleck}, J.~J., {Boily}, C.~M., {Lan{\c{c}}on}, A., \& {Deiters}, S. 2006,
  \mnras, 369, 1392, \dodoi{10.1111/j.1365-2966.2006.10390.x}

\bibitem[{Gagné {et~al.}(2021)Gagné, Faherty, Moranta, \&
  Popinchalk}]{gagne_number_2021}
Gagné, J., Faherty, J.~K., Moranta, L., \& Popinchalk, M. 2021, The
  Astrophysical Journal Letters, 915, L29, \dodoi{10.3847/2041-8213/ac0e9a}

\bibitem[{{Gaia Collaboration} {et~al.}(2022){Gaia Collaboration}, Vallenari,
  Brown, Prusti, \& {et al.}}]{gaia_collaboration_gaia_2022}
{Gaia Collaboration}, Vallenari, A., Brown, A., Prusti, T., \& {et al.} 2022,
  Astronomy \& Astrophysics, \dodoi{10.1051/0004-6361/202243940}

\bibitem[{{Gaia Collaboration} {et~al.}(2018{\natexlab{a}}){Gaia
  Collaboration}, {Katz}, {Antoja}, {Romero-G{\'o}mez}, {Drimmel}, {Reyl{\'e}},
  {Seabroke}, {Soubiran}, {Babusiaux}, {Di Matteo}, {Figueras}, {Poggio},
  {Robin}, {Evans}, {Brown}, {Vallenari}, {Prusti}, {de Bruijne},
  {Bailer-Jones}, {Biermann}, {Eyer}, {Jansen}, {Jordi}, {Klioner}, {Lammers},
  {Lindegren}, {Luri}, {Mignard}, {Panem}, {Pourbaix}, {Randich}, {Sartoretti},
  {Siddiqui}, {van Leeuwen}, {Walton}, {Arenou}, {Bastian}, {Cropper},
  {Lattanzi}, {Bakker}, {Cacciari}, {Casta n}, {Chaoul}, {Cheek}, {De Angeli},
  {Fabricius}, {Guerra}, {Holl}, {Masana}, {Messineo}, {Mowlavi},
  {Nienartowicz}, {Panuzzo}, {Portell}, {Riello}, {Tanga}, {Th{\'e}venin},
  {Gracia-Abril}, {Comoretto}, {Garcia-Reinaldos}, {Teyssier}, {Altmann},
  {Andrae}, {Audard}, {Bellas-Velidis}, {Benson}, {Berthier}, {Blomme},
  {Burgess}, {Busso}, {Carry}, {Cellino}, {Clementini}, {Clotet}, {Creevey},
  {Davidson}, {De Ridder}, {Delchambre}, {Dell'Oro}, {Ducourant},
  {Fern{\'a}ndez-Hern{\'a}ndez}, {Fouesneau}, {Fr{\'e}mat}, {Galluccio},
  {Garc{\'\i}a-Torres}, {Gonz{\'a}lez-N{\'u}{\~n}ez}, {Gonz{\'a}lez-Vidal},
  {Gosset}, {Guy}, {Halbwachs}, {Hambly}, {Harrison}, {Hern{\'a}ndez},
  {Hestroffer}, {Hodgkin}, {Hutton}, {Jasniewicz}, {Jean-Antoine-Piccolo},
  {Jordan}, {Korn}, {Krone-Martins}, {Lanzafame}, {Lebzelter}, {L{\"o}ffler},
  {Manteiga}, {Marrese}, {Mart{\'\i}n-Fleitas}, {Moitinho}, {Mora}, {Muinonen},
  {Osinde}, {Pancino}, {Pauwels}, {Petit}, {Recio-Blanco}, {Richards},
  {Rimoldini}, {Sarro}, {Siopis}, {Smith}, {Sozzetti}, {S{\"u}veges}, {Torra},
  {van Reeven}, {Abbas}, {Abreu Aramburu}, {Accart}, {Aerts}, {Altavilla},
  {{\'A}lvarez}, {Alvarez}, {Alves}, {Anderson}, {Andrei}, {Anglada Varela},
  {Antiche}, {Arcay}, {Astraatmadja}, {Bach}, {Baker},
  {Balaguer-N{\'u}{\~n}ez}, {Balm}, {Barache}, {Barata}, {Barbato}, {Barblan},
  {Barklem}, {Barrado}, {Barros}, {Barstow}, {Bartholom{\'e} Mu{\~n}oz},
  {Bassilana}, {Becciani}, {Bellazzini}, {Berihuete}, {Bertone}, {Bianchi},
  {Bienaym{\'e}}, {Blanco-Cuaresma}, {Boch}, {Boeche}, {Bombrun}, {Borrachero},
  {Bossini}, {Bouquillon}, {Bourda}, {Bragaglia}, {Bramante}, {Breddels},
  {Bressan}, {Brouillet}, {Br{\"u}semeister}, {Brugaletta}, {Bucciarelli},
  {Burlacu}, {Busonero}, {Butkevich}, {Buzzi}, {Caffau}, {Cancelliere},
  {Cannizzaro}, {Cantat-Gaudin}, {Carballo}, {Carlucci}, {Carrasco},
  {Casamiquela}, {Castellani}, {Castro-Ginard}, {Charlot}, {Chemin},
  {Chiavassa}, {Cocozza}, {Costigan}, {Cowell}, {Crifo}, {Crosta}, {Crowley},
  {Cuypers}, {Dafonte}, {Damerdji}, {Dapergolas}, {David}, {David}, {de
  Laverny}, {De Luise}, {De March}, {de Souza}, {de Torres}, {Debosscher}, {del
  Pozo}, {Delbo}, {Delgado}, {Delgado}, {Diakite}, {Diener}, {Distefano},
  {Dolding}, {Drazinos}, {Dur{\'a}n}, {Edvardsson}, {Enke}, {Eriksson},
  {Esquej}, {Eynard Bontemps}, {Fabre}, {Fabrizio}, {Faigler}, {Falc a},
  {Farr{\`a}s Casas}, {Federici}, {Fedorets}, {Fernique}, {Filippi},
  {Findeisen}, {Fonti}, {Fraile}, {Fraser}, {Fr{\'e}zouls}, {Gai}, {Galleti},
  {Garabato}, {Garc{\'\i}a-Sedano}, {Garofalo}, {Garralda}, {Gavel}, {Gavras},
  {Gerssen}, {Geyer}, {Giacobbe}, {Gilmore}, {Girona}, {Giuffrida}, {Glass},
  {Gomes}, {Granvik}, {Gueguen}, {Guerrier}, {Guiraud}, {Guti{\'e}}, {Haigron},
  {Hatzidimitriou}, {Hauser}, {Haywood}, {Heiter}, {Helmi}, {Heu}, {Hilger},
  {Hobbs}, {Hofmann}, {Holland}, {Huckle}, {Hypki}, {Icardi}, {Jan{\ss}en},
  {Jevardat de Fombelle}, {Jonker}, {Juh{\'a}sz}, {Julbe}, {Karampelas},
  {Kewley}, {Klar}, {Kochoska}, {Kohley}, {Kolenberg}, {Kontizas}, {Kontizas},
  {Koposov}, {Kordopatis}, {Kostrzewa-Rutkowska}, {Koubsky}, {Lambert},
  {Lanza}, {Lasne}, {Lavigne}, {Le Fustec}, {Le Poncin-Lafitte}, {Lebreton},
  {Leccia}, {Leclerc}, {Lecoeur-Taibi}, {Lenhardt}, {Leroux}, {Liao}, {Licata},
  {Lindstr{\o}m}, {Lister}, {Livanou}, {Lobel}, {L{\'o}pez}, {Managau}, {Mann},
  {Mantelet}, {Marchal}, {Marchant}, {Marconi}, {Marinoni}, {Marschalk{\'o}},
  {Marshall}, {Martino}, {Marton}, {Mary}, {Massari}, {Matijevi{\v{c}}},
  {Mazeh}, {McMillan}, {Messina}, {Michalik}, {Millar}, {Molina}, {Molinaro},
  {Moln{\'a}r}, {Montegriffo}, {Mor}, {Morbidelli}, {Morel}, {Morris},
  {Mulone}, {Muraveva}, {Musella}, {Nelemans}, {Nicastro}, {Noval},
  {O'Mullane}, {Ord{\'e}novic}, {Ord{\'o}{\~n}ez-Blanco}, {Osborne}, {Pagani},
  {Pagano}, {Pailler}, {Palacin}, {Palaversa}, {Panahi}, {Pawlak},
  {Piersimoni}, {Pineau}, {Plachy}, {Plum}, {Poujoulet}, {Pr{\v{s}}a},
  {Pulone}, {Racero}, {Ragaini}, {Rambaux}, {Ramos-Lerate}, {Regibo}, {Riclet},
  {Ripepi}, {Riva}, {Rivard}, {Rixon}, {Roegiers}, {Roelens}, {Rowell},
  {Royer}, {Ruiz-Dern}, {Sadowski}, {Sagrist{\`a} Sell{\'e}s}, {Sahlmann},
  {Salgado}, {Salguero}, {Sanna}, {Santana-Ros}, {Sarasso}, {Savietto},
  {Schultheis}, {Sciacca}, {Segol}, {Segovia}, {S{\'e}gransan}, {Shih},
  {Siltala}, {Silva}, {Smart}, {Smith}, {Solano}, {Solitro}, {Sordo}, {Soria
  Nieto}, {Souchay}, {Spagna}, {Spoto}, {Stampa}, {Steele},
  {Steidelm{\"u}ller}, {Stephenson}, {Stoev}, {Suess}, {Surdej}, {Szabados},
  {Szegedi-Elek}, {Tapiador}, {Taris}, {Tauran}, {Taylor}, {Teixeira},
  {Terrett}, {Teyssandier}, {Thuillot}, {Titarenko}, {Torra Clotet}, {Turon},
  {Ulla}, {Utrilla}, {Uzzi}, {Vaillant}, {Valentini}, {Valette}, {van Elteren},
  {Van Hemelryck}, {van Leeuwen}, {Vaschetto}, {Vecchiato}, {Veljanoski},
  {Viala}, {Vicente}, {Vogt}, {von Essen}, {Voss}, {Votruba}, {Voutsinas},
  {Walmsley}, {Weiler}, {Wertz}, {Wevers}, {Wyrzykowski}, {Yoldas},
  {{\v{Z}}erjal}, {Ziaeepour}, {Zorec}, {Zschocke}, {Zucker}, {Zurbach}, \&
  {Zwitter}}]{gaia2018kin}
{Gaia Collaboration}, {Katz}, D., {Antoja}, T., {et~al.} 2018{\natexlab{a}},
  \aap, 616, A11, \dodoi{10.1051/0004-6361/201832865}

\bibitem[{{Gaia Collaboration} {et~al.}(2018{\natexlab{b}}){Gaia
  Collaboration}, {Brown}, {Vallenari}, {Prusti}, {de Bruijne}, {Babusiaux},
  {Bailer-Jones}, {Biermann}, {Evans}, {Eyer}, {Jansen}, {Jordi}, {Klioner},
  {Lammers}, {Lindegren}, {Luri}, {Mignard}, {Panem}, {Pourbaix}, {Randich},
  {Sartoretti}, {Siddiqui}, {Soubiran}, {van Leeuwen}, {Walton}, {Arenou},
  {Bastian}, {Cropper}, {Drimmel}, {Katz}, {Lattanzi}, {Bakker}, {Cacciari},
  {Casta{\~n}eda}, {Chaoul}, {Cheek}, {De Angeli}, {Fabricius}, {Guerra},
  {Holl}, {Masana}, {Messineo}, {Mowlavi}, {Nienartowicz}, {Panuzzo},
  {Portell}, {Riello}, {Seabroke}, {Tanga}, {Th{\'e}venin}, {Gracia-Abril},
  {Comoretto}, {Garcia-Reinaldos}, {Teyssier}, {Altmann}, {Andrae}, {Audard},
  {Bellas-Velidis}, {Benson}, {Berthier}, {Blomme}, {Burgess}, {Busso},
  {Carry}, {Cellino}, {Clementini}, {Clotet}, {Creevey}, {Davidson}, {De
  Ridder}, {Delchambre}, {Dell'Oro}, {Ducourant},
  {Fern{\'a}ndez-Hern{\'a}ndez}, {Fouesneau}, {Fr{\'e}mat}, {Galluccio},
  {Garc{\'\i}a-Torres}, {Gonz{\'a}lez-N{\'u}{\~n}ez}, {Gonz{\'a}lez-Vidal},
  {Gosset}, {Guy}, {Halbwachs}, {Hambly}, {Harrison}, {Hern{\'a}ndez},
  {Hestroffer}, {Hodgkin}, {Hutton}, {Jasniewicz}, {Jean-Antoine-Piccolo},
  {Jordan}, {Korn}, {Krone-Martins}, {Lanzafame}, {Lebzelter}, {L{\"o}ffler},
  {Manteiga}, {Marrese}, {Mart{\'\i}n-Fleitas}, {Moitinho}, {Mora}, {Muinonen},
  {Osinde}, {Pancino}, {Pauwels}, {Petit}, {Recio-Blanco}, {Richards},
  {Rimoldini}, {Robin}, {Sarro}, {Siopis}, {Smith}, {Sozzetti}, {S{\"u}veges},
  {Torra}, {van Reeven}, {Abbas}, {Abreu Aramburu}, {Accart}, {Aerts},
  {Altavilla}, {{\'A}lvarez}, {Alvarez}, {Alves}, {Anderson}, {Andrei},
  {Anglada Varela}, {Antiche}, {Antoja}, {Arcay}, {Astraatmadja}, {Bach},
  {Baker}, {Balaguer-N{\'u}{\~n}ez}, {Balm}, {Barache}, {Barata}, {Barbato},
  {Barblan}, {Barklem}, {Barrado}, {Barros}, {Barstow}, {Bartholom{\'e}
  Mu{\~n}oz}, {Bassilana}, {Becciani}, {Bellazzini}, {Berihuete}, {Bertone},
  {Bianchi}, {Bienaym{\'e}}, {Blanco-Cuaresma}, {Boch}, {Boeche}, {Bombrun},
  {Borrachero}, {Bossini}, {Bouquillon}, {Bourda}, {Bragaglia}, {Bramante},
  {Breddels}, {Bressan}, {Brouillet}, {Br{\"u}semeister}, {Brugaletta},
  {Bucciarelli}, {Burlacu}, {Busonero}, {Butkevich}, {Buzzi}, {Caffau},
  {Cancelliere}, {Cannizzaro}, {Cantat-Gaudin}, {Carballo}, {Carlucci},
  {Carrasco}, {Casamiquela}, {Castellani}, {Castro-Ginard}, {Charlot},
  {Chemin}, {Chiavassa}, {Cocozza}, {Costigan}, {Cowell}, {Crifo}, {Crosta},
  {Crowley}, {Cuypers}, {Dafonte}, {Damerdji}, {Dapergolas}, {David}, {David},
  {de Laverny}, {De Luise}, {De March}, {de Martino}, {de Souza}, {de Torres},
  {Debosscher}, {del Pozo}, {Delbo}, {Delgado}, {Delgado}, {Di Matteo},
  {Diakite}, {Diener}, {Distefano}, {Dolding}, {Drazinos}, {Dur{\'a}n},
  {Edvardsson}, {Enke}, {Eriksson}, {Esquej}, {Eynard Bontemps}, {Fabre},
  {Fabrizio}, {Faigler}, {Falc{\~a}o}, {Farr{\`a}s Casas}, {Federici},
  {Fedorets}, {Fernique}, {Figueras}, {Filippi}, {Findeisen}, {Fonti},
  {Fraile}, {Fraser}, {Fr{\'e}zouls}, {Gai}, {Galleti}, {Garabato},
  {Garc{\'\i}a-Sedano}, {Garofalo}, {Garralda}, {Gavel}, {Gavras}, {Gerssen},
  {Geyer}, {Giacobbe}, {Gilmore}, {Girona}, {Giuffrida}, {Glass}, {Gomes},
  {Granvik}, {Gueguen}, {Guerrier}, {Guiraud}, {Guti{\'e}rrez-S{\'a}nchez},
  {Haigron}, {Hatzidimitriou}, {Hauser}, {Haywood}, {Heiter}, {Helmi}, {Heu},
  {Hilger}, {Hobbs}, {Hofmann}, {Holland}, {Huckle}, {Hypki}, {Icardi},
  {Jan{\ss}en}, {Jevardat de Fombelle}, {Jonker}, {Juh{\'a}sz}, {Julbe},
  {Karampelas}, {Kewley}, {Klar}, {Kochoska}, {Kohley}, {Kolenberg},
  {Kontizas}, {Kontizas}, {Koposov}, {Kordopatis}, {Kostrzewa-Rutkowska},
  {Koubsky}, {Lambert}, {Lanza}, {Lasne}, {Lavigne}, {Le Fustec}, {Le
  Poncin-Lafitte}, {Lebreton}, {Leccia}, {Leclerc}, {Lecoeur-Taibi},
  {Lenhardt}, {Leroux}, {Liao}, {Licata}, {Lindstr{\o}m}, {Lister}, {Livanou},
  {Lobel}, {L{\'o}pez}, {Managau}, {Mann}, {Mantelet}, {Marchal}, {Marchant},
  {Marconi}, {Marinoni}, {Marschalk{\'o}}, {Marshall}, {Martino}, {Marton},
  {Mary}, {Massari}, {Matijevi{\v{c}}}, {Mazeh}, {McMillan}, {Messina},
  {Michalik}, {Millar}, {Molina}, {Molinaro}, {Moln{\'a}r}, {Montegriffo},
  {Mor}, {Morbidelli}, {Morel}, {Morris}, {Mulone}, {Muraveva}, {Musella},
  {Nelemans}, {Nicastro}, {Noval}, {O'Mullane}, {Ord{\'e}novic},
  {Ord{\'o}{\~n}ez-Blanco}, {Osborne}, {Pagani}, {Pagano}, {Pailler},
  {Palacin}, {Palaversa}, {Panahi}, {Pawlak}, {Piersimoni}, {Pineau}, {Plachy},
  {Plum}, {Poggio}, {Poujoulet}, {Pr{\v{s}}a}, {Pulone}, {Racero}, {Ragaini},
  {Rambaux}, {Ramos-Lerate}, {Regibo}, {Reyl{\'e}}, {Riclet}, {Ripepi}, {Riva},
  {Rivard}, {Rixon}, {Roegiers}, {Roelens}, {Romero-G{\'o}mez}, {Rowell},
  {Royer}, {Ruiz-Dern}, {Sadowski}, {Sagrist{\`a} Sell{\'e}s}, {Sahlmann},
  {Salgado}, {Salguero}, {Sanna}, {Santana-Ros}, {Sarasso}, {Savietto},
  {Schultheis}, {Sciacca}, {Segol}, {Segovia}, {S{\'e}gransan}, {Shih},
  {Siltala}, {Silva}, {Smart}, {Smith}, {Solano}, {Solitro}, {Sordo}, {Soria
  Nieto}, {Souchay}, {Spagna}, {Spoto}, {Stampa}, {Steele},
  {Steidelm{\"u}ller}, {Stephenson}, {Stoev}, {Suess}, {Surdej}, {Szabados},
  {Szegedi-Elek}, {Tapiador}, {Taris}, {Tauran}, {Taylor}, {Teixeira},
  {Terrett}, {Teyssandier}, {Thuillot}, {Titarenko}, {Torra Clotet}, {Turon},
  {Ulla}, {Utrilla}, {Uzzi}, {Vaillant}, {Valentini}, {Valette}, {van Elteren},
  {Van Hemelryck}, {van Leeuwen}, {Vaschetto}, {Vecchiato}, {Veljanoski},
  {Viala}, {Vicente}, {Vogt}, {von Essen}, {Voss}, {Votruba}, {Voutsinas},
  {Walmsley}, {Weiler}, {Wertz}, {Wevers}, {Wyrzykowski}, {Yoldas},
  {{\v{Z}}erjal}, {Ziaeepour}, {Zorec}, {Zschocke}, {Zucker}, {Zurbach}, \&
  {Zwitter}}]{gaia2018}
{Gaia Collaboration}, {Brown}, A.~G.~A., {Vallenari}, A., {et~al.}
  2018{\natexlab{b}}, \aap, 616, A1, \dodoi{10.1051/0004-6361/201833051}

\bibitem[{{Gaia Collaboration} {et~al.}(2021){Gaia Collaboration}, {Brown},
  {Vallenari}, {Prusti}, {de Bruijne}, {Babusiaux}, {Biermann}, {Creevey},
  {Evans}, {Eyer}, {Hutton}, {Jansen}, {Jordi}, {Klioner}, {Lammers},
  {Lindegren}, {Luri}, {Mignard}, {Panem}, {Pourbaix}, {Randich}, {Sartoretti},
  {Soubiran}, {Walton}, {Arenou}, {Bailer-Jones}, {Bastian}, {Cropper},
  {Drimmel}, {Katz}, {Lattanzi}, {van Leeuwen}, {Bakker}, {Cacciari},
  {Casta{\~n}eda}, {De Angeli}, {Ducourant}, {Fabricius}, {Fouesneau},
  {Fr{\'e}mat}, {Guerra}, {Guerrier}, {Guiraud}, {Jean-Antoine Piccolo},
  {Masana}, {Messineo}, {Mowlavi}, {Nicolas}, {Nienartowicz}, {Pailler},
  {Panuzzo}, {Riclet}, {Roux}, {Seabroke}, {Sordo}, {Tanga}, {Th{\'e}venin},
  {Gracia-Abril}, {Portell}, {Teyssier}, {Altmann}, {Andrae}, {Bellas-Velidis},
  {Benson}, {Berthier}, {Blomme}, {Brugaletta}, {Burgess}, {Busso}, {Carry},
  {Cellino}, {Cheek}, {Clementini}, {Damerdji}, {Davidson}, {Delchambre},
  {Dell'Oro}, {Fern{\'a}ndez-Hern{\'a}ndez}, {Galluccio}, {Garc{\'\i}a-Lario},
  {Garcia-Reinaldos}, {Gonz{\'a}lez-N{\'u}{\~n}ez}, {Gosset}, {Haigron},
  {Halbwachs}, {Hambly}, {Harrison}, {Hatzidimitriou}, {Heiter},
  {Hern{\'a}ndez}, {Hestroffer}, {Hodgkin}, {Holl}, {Jan{\ss}en}, {Jevardat de
  Fombelle}, {Jordan}, {Krone-Martins}, {Lanzafame}, {L{\"o}ffler}, {Lorca},
  {Manteiga}, {Marchal}, {Marrese}, {Moitinho}, {Mora}, {Muinonen}, {Osborne},
  {Pancino}, {Pauwels}, {Petit}, {Recio-Blanco}, {Richards}, {Riello},
  {Rimoldini}, {Robin}, {Roegiers}, {Rybizki}, {Sarro}, {Siopis}, {Smith},
  {Sozzetti}, {Ulla}, {Utrilla}, {van Leeuwen}, {van Reeven}, {Abbas}, {Abreu
  Aramburu}, {Accart}, {Aerts}, {Aguado}, {Ajaj}, {Altavilla}, {{\'A}lvarez},
  {{\'A}lvarez Cid-Fuentes}, {Alves}, {Anderson}, {Anglada Varela}, {Antoja},
  {Audard}, {Baines}, {Baker}, {Balaguer-N{\'u}{\~n}ez}, {Balbinot}, {Balog},
  {Barache}, {Barbato}, {Barros}, {Barstow}, {Bartolom{\'e}}, {Bassilana},
  {Bauchet}, {Baudesson-Stella}, {Becciani}, {Bellazzini}, {Bernet}, {Bertone},
  {Bianchi}, {Blanco-Cuaresma}, {Boch}, {Bombrun}, {Bossini}, {Bouquillon},
  {Bragaglia}, {Bramante}, {Breedt}, {Bressan}, {Brouillet}, {Bucciarelli},
  {Burlacu}, {Busonero}, {Butkevich}, {Buzzi}, {Caffau}, {Cancelliere},
  {C{\'a}novas}, {Cantat-Gaudin}, {Carballo}, {Carlucci}, {Carnerero},
  {Carrasco}, {Casamiquela}, {Castellani}, {Castro-Ginard}, {Castro Sampol},
  {Chaoul}, {Charlot}, {Chemin}, {Chiavassa}, {Cioni}, {Comoretto}, {Cooper},
  {Cornez}, {Cowell}, {Crifo}, {Crosta}, {Crowley}, {Dafonte}, {Dapergolas},
  {David}, {David}, {de Laverny}, {De Luise}, {De March}, {De Ridder}, {de
  Souza}, {de Teodoro}, {de Torres}, {del Peloso}, {del Pozo}, {Delbo},
  {Delgado}, {Delgado}, {Delisle}, {Di Matteo}, {Diakite}, {Diener},
  {Distefano}, {Dolding}, {Eappachen}, {Edvardsson}, {Enke}, {Esquej}, {Fabre},
  {Fabrizio}, {Faigler}, {Fedorets}, {Fernique}, {Fienga}, {Figueras},
  {Fouron}, {Fragkoudi}, {Fraile}, {Franke}, {Gai}, {Garabato},
  {Garcia-Gutierrez}, {Garc{\'\i}a-Torres}, {Garofalo}, {Gavras}, {Gerlach},
  {Geyer}, {Giacobbe}, {Gilmore}, {Girona}, {Giuffrida}, {Gomel}, {Gomez},
  {Gonzalez-Santamaria}, {Gonz{\'a}lez-Vidal}, {Granvik},
  {Guti{\'e}rrez-S{\'a}nchez}, {Guy}, {Hauser}, {Haywood}, {Helmi}, {Hidalgo},
  {Hilger}, {H{\l}adczuk}, {Hobbs}, {Holland}, {Huckle}, {Jasniewicz},
  {Jonker}, {Juaristi Campillo}, {Julbe}, {Karbevska}, {Kervella}, {Khanna},
  {Kochoska}, {Kontizas}, {Kordopatis}, {Korn}, {Kostrzewa-Rutkowska},
  {Kruszy{\'n}ska}, {Lambert}, {Lanza}, {Lasne}, {Le Campion}, {Le Fustec},
  {Lebreton}, {Lebzelter}, {Leccia}, {Leclerc}, {Lecoeur-Taibi}, {Liao},
  {Licata}, {Lindstr{\o}m}, {Lister}, {Livanou}, {Lobel}, {Madrero Pardo},
  {Managau}, {Mann}, {Marchant}, {Marconi}, {Marcos Santos}, {Marinoni},
  {Marocco}, {Marshall}, {Martin Polo}, {Mart{\'\i}n-Fleitas}, {Masip},
  {Massari}, {Mastrobuono-Battisti}, {Mazeh}, {McMillan}, {Messina},
  {Michalik}, {Millar}, {Mints}, {Molina}, {Molinaro}, {Moln{\'a}r},
  {Montegriffo}, {Mor}, {Morbidelli}, {Morel}, {Morris}, {Mulone}, {Munoz},
  {Muraveva}, {Murphy}, {Musella}, {Noval}, {Ord{\'e}novic}, {Orr{\`u}},
  {Osinde}, {Pagani}, {Pagano}, {Palaversa}, {Palicio}, {Panahi}, {Pawlak},
  {Pe{\~n}alosa Esteller}, {Penttil{\"a}}, {Piersimoni}, {Pineau}, {Plachy},
  {Plum}, {Poggio}, {Poretti}, {Poujoulet}, {Pr{\v{s}}a}, {Pulone}, {Racero},
  {Ragaini}, {Rainer}, {Raiteri}, {Rambaux}, {Ramos}, {Ramos-Lerate}, {Re
  Fiorentin}, {Regibo}, {Reyl{\'e}}, {Ripepi}, {Riva}, {Rixon}, {Robichon},
  {Robin}, {Roelens}, {Rohrbasser}, {Romero-G{\'o}mez}, {Rowell}, {Royer},
  {Rybicki}, {Sadowski}, {Sagrist{\`a} Sell{\'e}s}, {Sahlmann}, {Salgado},
  {Salguero}, {Samaras}, {Sanchez Gimenez}, {Sanna}, {Santove{\~n}a},
  {Sarasso}, {Schultheis}, {Sciacca}, {Segol}, {Segovia}, {S{\'e}gransan},
  {Semeux}, {Shahaf}, {Siddiqui}, {Siebert}, {Siltala}, {Slezak}, {Smart},
  {Solano}, {Solitro}, {Souami}, {Souchay}, {Spagna}, {Spoto}, {Steele},
  {Steidelm{\"u}ller}, {Stephenson}, {S{\"u}veges}, {Szabados}, {Szegedi-Elek},
  {Taris}, {Tauran}, {Taylor}, {Teixeira}, {Thuillot}, {Tonello}, {Torra},
  {Torra}, {Turon}, {Unger}, {Vaillant}, {van Dillen}, {Vanel}, {Vecchiato},
  {Viala}, {Vicente}, {Voutsinas}, {Weiler}, {Wevers}, {Wyrzykowski}, {Yoldas},
  {Yvard}, {Zhao}, {Zorec}, {Zucker}, {Zurbach}, \& {Zwitter}}]{gaia2021}
---. 2021, \aap, 649, A1, \dodoi{10.1051/0004-6361/202039657}

\bibitem[{{Green} {et~al.}(2019){Green}, {Schlafly}, {Zucker}, {Speagle}, \&
  {Finkbeiner}}]{green2019}
{Green}, G.~M., {Schlafly}, E., {Zucker}, C., {Speagle}, J.~S., \&
  {Finkbeiner}, D. 2019, \apj, 887, 93, \dodoi{10.3847/1538-4357/ab5362}

\bibitem[{He {et~al.}(2022)He, Wang, Luo, Li, Liu, \& Jiang}]{he_blind_2022}
He, Z., Wang, K., Luo, Y., {et~al.} 2022, A {Blind} {All}-sky {Search} for
  {Star} {Clusters} in {Gaia} {EDR3}: 886 {Clusters} within 1.2 kpc of the
  {Sun},  arXiv.
\newblock \url{http://arxiv.org/abs/2206.12170}

\bibitem[{{Hou} \& {Gao}(2014)}]{hou2014}
{Hou}, L.~G., \& {Gao}, X.~Y. 2014, \mnras, 438, 426,
  \dodoi{10.1093/mnras/stt2212}

\bibitem[{{Jerabkova} {et~al.}(2019){Jerabkova}, {Boffin}, {Beccari}, \&
  {Anderson}}]{jerabkova2019}
{Jerabkova}, T., {Boffin}, H. M.~J., {Beccari}, G., \& {Anderson}, R.~I. 2019,
  \mnras, 489, 4418, \dodoi{10.1093/mnras/stz2315}

\bibitem[{Katz {et~al.}(2022)Katz, Sartoretti, Guerrier, Panuzzo, Seabroke,
  Thévenin, Cropper, Benson, Blomme, Haigron, Marchal, Smith, Baker, Chemin,
  Damerdji, David, Dolding, Frémat, Gosset, Janßen, Jasniewicz, Lobel, Plum,
  Samaras, Snaith, Soubiran, Vanel, Zwitter, Antoja, Arenou, Babusiaux,
  Brouillet, Caffau, Di~Matteo, Fabre, Fabricius, Frakgoudi, Haywood, Huckle,
  Hottier, Lasne, Leclerc, Mastrobuono-Battisti, Royer, Teyssier, Zorec, Crifo,
  Piccolo, Turon, \& Viala}]{katz_gaia_2022}
Katz, D., Sartoretti, P., Guerrier, A., {et~al.} 2022, Gaia {Data} {Release} 3
  {Properties} and validation of the radial velocities,  arXiv.
\newblock \url{http://arxiv.org/abs/2206.05902}

\bibitem[{{Kounkel} \& {Covey}(2019)}]{kounkel2019}
{Kounkel}, M., \& {Covey}, K. 2019, \aj, 158, 122,
  \dodoi{10.3847/1538-3881/ab339a}

\bibitem[{Kushniruk {et~al.}(2020)Kushniruk, Bensby, Feltzing, Sahlholdt,
  Feuillet, \& Casagrande}]{kushniruk_hr_2020}
Kushniruk, I., Bensby, T., Feltzing, S., {et~al.} 2020, Astronomy \&
  Astrophysics, 638, A154, \dodoi{10.1051/0004-6361/202037923}

\bibitem[{{Kushniruk} {et~al.}(2017){Kushniruk}, {Schirmer}, \&
  {Bensby}}]{kushniruk2017}
{Kushniruk}, I., {Schirmer}, T., \& {Bensby}, T. 2017, \aap, 608, A73,
  \dodoi{10.1051/0004-6361/201731147}

\bibitem[{Lee {et~al.}(2022)Lee, Song, \& Murphy}]{lee_low-mass_2022}
Lee, J., Song, I., \& Murphy, S.~J. 2022, Monthly Notices of the Royal
  Astronomical Society, 511, 6179, \dodoi{10.1093/mnras/stac358}

\bibitem[{{Li} \& {Widrow}(2021)}]{liw2021}
{Li}, H., \& {Widrow}, L.~M. 2021, \mnras, 503, 1586,
  \dodoi{10.1093/mnras/stab574}

\bibitem[{{Li} {et~al.}(2021){Li}, {Pang}, \& {Tang}}]{li2021}
{Li}, Y., {Pang}, X., \& {Tang}, S.-Y. 2021, Research Notes of the American
  Astronomical Society, 5, 173, \dodoi{10.3847/2515-5172/ac1688}

\bibitem[{{Li}(2021)}]{liz2021}
{Li}, Z.-Y. 2021, \apj, 911, 107, \dodoi{10.3847/1538-4357/abea17}

\bibitem[{{Li} \& {Shen}(2020)}]{li2020}
{Li}, Z.-Y., \& {Shen}, J. 2020, \apj, 890, 85,
  \dodoi{10.3847/1538-4357/ab6b21}

\bibitem[{{Liu} \& {Pang}(2019)}]{liu2019}
{Liu}, L., \& {Pang}, X. 2019, \apjs, 245, 32, \dodoi{10.3847/1538-4365/ab530a}

\bibitem[{{McMillan} {et~al.}(2007){McMillan}, {Vesperini}, \& {Portegies
  Zwart}}]{mcmillan2007}
{McMillan}, S. L.~W., {Vesperini}, E., \& {Portegies Zwart}, S.~F. 2007, \apjl,
  655, L45, \dodoi{10.1086/511763}

\bibitem[{Messina {et~al.}(2022)Messina, Nardiello, Desidera, Baratella,
  Benatti, Biazzo, \& D’Orazi}]{messina_gyrochronological_2022}
Messina, S., Nardiello, D., Desidera, S., {et~al.} 2022, Astronomy \&
  Astrophysics, 657, L3, \dodoi{10.1051/0004-6361/202142276}

\bibitem[{{Millman} \& {Aivazis}(2011)}]{millman2011}
{Millman}, K.~J., \& {Aivazis}, M. 2011, Computing in Science and Engineering,
  13, 9, \dodoi{10.1109/MCSE.2011.36}

\bibitem[{Minchev {et~al.}(2010)Minchev, Boily, Siebert, \&
  Bienayme}]{minchev_low-velocity_2010}
Minchev, I., Boily, C., Siebert, A., \& Bienayme, O. 2010, Monthly Notices of
  the Royal Astronomical Society, 407, 2122,
  \dodoi{10.1111/j.1365-2966.2010.17060.x}

\bibitem[{Miret-Roig {et~al.}(2020)Miret-Roig, Galli, Brandner, Bouy, Barrado,
  Olivares, Antoja, Romero-Gómez, Figueras, \&
  Lillo-Box}]{miret-roig_dynamical_2020}
Miret-Roig, N., Galli, P. A.~B., Brandner, W., {et~al.} 2020, Astronomy \&
  Astrophysics, 642, A179, \dodoi{10.1051/0004-6361/202038765}

\bibitem[{{Miville-Desch{\^e}nes} \& {Lagache}(2005)}]{miville2005}
{Miville-Desch{\^e}nes}, M.-A., \& {Lagache}, G. 2005, \apjs, 157, 302,
  \dodoi{10.1086/427938}

\bibitem[{{Pang} {et~al.}(2013){Pang}, {Grebel}, {Allison}, {Goodwin},
  {Altmann}, {Harbeck}, {Moffat}, \& {Drissen}}]{pang2013}
{Pang}, X., {Grebel}, E.~K., {Allison}, R.~J., {et~al.} 2013, \apj, 764, 73,
  \dodoi{10.1088/0004-637X/764/1/73}

\bibitem[{{Pang} {et~al.}(2020){Pang}, {Li}, {Tang}, {Pasquato}, \&
  {Kouwenhoven}}]{pang2020}
{Pang}, X., {Li}, Y., {Tang}, S.-Y., {Pasquato}, M., \& {Kouwenhoven}, M.~B.~N.
  2020, \apjl, 900, L4, \dodoi{10.3847/2041-8213/abad28}

\bibitem[{{Pang} {et~al.}(2021{\natexlab{a}}){Pang}, {Li}, {Yu}, {Tang},
  {Dinnbier}, {Kroupa}, {Pasquato}, \& {Kouwenhoven}}]{pang2021a}
{Pang}, X., {Li}, Y., {Yu}, Z., {et~al.} 2021{\natexlab{a}}, \apj, 912, 162,
  \dodoi{10.3847/1538-4357/abeaac}

\bibitem[{{Pang} {et~al.}(2021{\natexlab{b}}){Pang}, {Yu}, {Tang}, {Hong},
  {Yuan}, {Pasquato}, \& {Kouwenhoven}}]{pang2021b}
{Pang}, X., {Yu}, Z., {Tang}, S.-Y., {et~al.} 2021{\natexlab{b}}, \apj, 923,
  20, \dodoi{10.3847/1538-4357/ac2838}

\bibitem[{{Pang} {et~al.}(2022){Pang}, {Tang}, {Li}, {Yu}, {Wang}, {Li}, {Li},
  {Wang}, {Wang}, {Zhang}, {Pasquato}, \& {Kouwenhoven}}]{pang2022}
{Pang}, X., {Tang}, S.-Y., {Li}, Y., {et~al.} 2022, \apj, 931, 156,
  \dodoi{10.3847/1538-4357/ac674e}

\bibitem[{{Pinfield} {et~al.}(1998){Pinfield}, {Jameson}, \&
  {Hodgkin}}]{pinfield1998}
{Pinfield}, D.~J., {Jameson}, R.~F., \& {Hodgkin}, S.~T. 1998, \mnras, 299,
  955, \dodoi{10.1046/j.1365-8711.1998.01754.x}

\bibitem[{Recio-Blanco {et~al.}(2022)Recio-Blanco, de~Laverny, Palicio,
  Kordopatis, Álvarez, Schultheis, Contursi, Zhao, Elipe, Ordenovic, Manteiga,
  Dafonte, Oreshina-Slezak, Bijaoui, Fremat, Seabroke, Pailler, Spitoni,
  Poggio, Creevey, Aramburu, Accart, Andrae, Bailer-Jones, Bellas-Velidis,
  Brouillet, Brugaletta, Burlacu, Carballo, Casamiquela, Chiavassa, Cooper,
  Dapergolas, Delchambre, Dharmawardena, Drimmel, Edvardsson, Fouesneau,
  Garabato, Garcia-Lario, Garcia-Torres, Gavel, Gomez, Gonzalez-Santamaria,
  Hatzidimitriou, Heiter, Piccolo, Kontizas, Korn, Lanzafame, Lebreton, Fustec,
  Licata, Lindstrom, Livanou, Lobel, Lorca, Romeo, Marocco, Marshall, Mary,
  Nicolas, Pallas-Quintela, Panem, Pichon, Riclet, Robin, Rybizki, Santovena,
  Silvelo, Smart, Sarro, Sordo, Soubiran, Suvege, Ulla, Vallenari, Zorec,
  Utrilla, \& Bakker}]{recio-blanco_gaia_2022}
Recio-Blanco, A., de~Laverny, P., Palicio, P.~A., {et~al.} 2022, Gaia {Data}
  {Release} 3: {Analysis} of {RVS} spectra using the {General} {Stellar}
  {Parametriser} from spectroscopy,  arXiv.
\newblock \url{http://arxiv.org/abs/2206.05541}

\bibitem[{{Reid} {et~al.}(2019){Reid}, {Menten}, {Brunthaler}, {Zheng}, {Dame},
  {Xu}, {Li}, {Sakai}, {Wu}, {Immer}, {Zhang}, {Sanna}, {Moscadelli}, {Rygl},
  {Bartkiewicz}, {Hu}, {Quiroga-Nu{\~n}ez}, \& {van Langevelde}}]{reid2019}
{Reid}, M.~J., {Menten}, K.~M., {Brunthaler}, A., {et~al.} 2019, \apj, 885,
  131, \dodoi{10.3847/1538-4357/ab4a11}

\bibitem[{{Riello} {et~al.}(2021){Riello}, {De Angeli}, {Evans}, {Montegriffo},
  {Carrasco}, {Busso}, {Palaversa}, {Burgess}, {Diener}, {Davidson}, {Rowell},
  {Fabricius}, {Jordi}, {Bellazzini}, {Pancino}, {Harrison}, {Cacciari}, {van
  Leeuwen}, {Hambly}, {Hodgkin}, {Osborne}, {Altavilla}, {Barstow}, {Brown},
  {Castellani}, {Cowell}, {De Luise}, {Gilmore}, {Giuffrida}, {Hidalgo},
  {Holland}, {Marinoni}, {Pagani}, {Piersimoni}, {Pulone}, {Ragaini}, {Rainer},
  {Richards}, {Sanna}, {Walton}, {Weiler}, \& {Yoldas}}]{riello2021}
{Riello}, M., {De Angeli}, F., {Evans}, D.~W., {et~al.} 2021, \aap, 649, A3,
  \dodoi{10.1051/0004-6361/202039587}

\bibitem[{{Rybizki} {et~al.}(2020){Rybizki}, {Demleitner}, {Bailer-Jones},
  {Tio}, {Cantat-Gaudin}, {Fouesneau}, {Chen}, {Andrae}, {Girardi}, \&
  {Sharma}}]{rybizki2020}
{Rybizki}, J., {Demleitner}, M., {Bailer-Jones}, C., {et~al.} 2020, \pasp, 132,
  074501, \dodoi{10.1088/1538-3873/ab8cb0}

\bibitem[{S.~De~Simone {et~al.}(2004)S.~De~Simone, Wu, \&
  Tremaine}]{s_de_simone_stellar_2004}
S.~De~Simone, R., Wu, X., \& Tremaine, S. 2004, Monthly Notices of the Royal
  Astronomical Society, 350, 627, \dodoi{10.1111/j.1365-2966.2004.07675.x}

\bibitem[{{Sellwood} \& {Binney}(2002)}]{sellwood2002}
{Sellwood}, J.~A., \& {Binney}, J.~J. 2002, \mnras, 336, 785,
  \dodoi{10.1046/j.1365-8711.2002.05806.x}

\bibitem[{Simpson {et~al.}(2012)Simpson, Povich, Kendrew, Lintott, Bressert,
  Arvidsson, Cyganowski, Maddison, Schawinski, Sherman, Smith, \&
  Wolf-Chase}]{simpson_milky_2012}
Simpson, R.~J., Povich, M.~S., Kendrew, S., {et~al.} 2012, Monthly Notices of
  the Royal Astronomical Society, 424, 2442,
  \dodoi{10.1111/j.1365-2966.2012.20770.x}

\bibitem[{Skuljan {et~al.}(1999)Skuljan, Hearnshaw, \&
  Cottrell}]{skuljan_velocity_1999}
Skuljan, J., Hearnshaw, J.~B., \& Cottrell, P.~L. 1999, Monthly Notices of the
  Royal Astronomical Society, 308, 731,
  \dodoi{10.1046/j.1365-8711.1999.02736.x}

\bibitem[{{Tang} {et~al.}(2019){Tang}, {Pang}, {Yuan}, {Chen}, {Hong},
  {Goldman}, {Just}, {Shukirgaliyev}, \& {Lin}}]{tang2019}
{Tang}, S.-Y., {Pang}, X., {Yuan}, Z., {et~al.} 2019, \apj, 877, 12,
  \dodoi{10.3847/1538-4357/ab13b0}

\bibitem[{{Taylor}(2005)}]{taylor2005}
{Taylor}, M.~B. 2005, in Astronomical Society of the Pacific Conference Series,
  Vol. 347, Astronomical Data Analysis Software and Systems XIV, ed.
  P.~{Shopbell}, M.~{Britton}, \& R.~{Ebert}, 29

\bibitem[{{Wang} \& {Jerabkova}(2021)}]{wang2021a}
{Wang}, L., \& {Jerabkova}, T. 2021, \aap, 655, A71,
  \dodoi{10.1051/0004-6361/202141838}

\bibitem[{Yan {et~al.}(2016)Yan, Xu, Zhang, Lu, Chen, \&
  Tang}]{yan_molecular_2016}
Yan, Q.-z., Xu, Y., Zhang, B., {et~al.} 2016, The Astronomical Journal, 152,
  117, \dodoi{10.3847/0004-6256/152/5/117}

\bibitem[{{Yuan} {et~al.}(2018){Yuan}, {Chang}, {Banerjee}, {Han}, {Kang}, \&
  {Smith}}]{yuan2018}
{Yuan}, Z., {Chang}, J., {Banerjee}, P., {et~al.} 2018, \apj, 863, 26,
  \dodoi{10.3847/1538-4357/aacd0d}

\bibitem[{{Yuan} {et~al.}(2020{\natexlab{a}}){Yuan}, {Chang}, {Beers}, \&
  {Huang}}]{yuan2020a}
{Yuan}, Z., {Chang}, J., {Beers}, T.~C., \& {Huang}, Y. 2020{\natexlab{a}},
  \apjl, 898, L37, \dodoi{10.3847/2041-8213/aba49f}

\bibitem[{{Yuan} {et~al.}(2020{\natexlab{b}}){Yuan}, {Myeong}, {Beers},
  {Evans}, {Lee}, {Banerjee}, {Gudin}, {Hattori}, {Li}, {Matsuno}, {Placco},
  {Smith}, {Whitten}, \& {Zhao}}]{yuan2020b}
{Yuan}, Z., {Myeong}, G.~C., {Beers}, T.~C., {et~al.} 2020{\natexlab{b}}, \apj,
  891, 39, \dodoi{10.3847/1538-4357/ab6ef7}

\bibitem[{{Zhang} {et~al.}(2019){Zhang}, {Zhao}, {Oswalt}, {Fang}, {Zhao},
  {Liang}, {Ye}, \& {Zhong}}]{Zhang2019}
{Zhang}, J., {Zhao}, J., {Oswalt}, T.~D., {et~al.} 2019, \apj, 887, 84,
  \dodoi{10.3847/1538-4357/ab4efe}

\bibitem[{{Zhao} {et~al.}(2009){Zhao}, {Zhao}, \& {Chen}}]{zhao2009}
{Zhao}, J., {Zhao}, G., \& {Chen}, Y. 2009, \apjl, 692, L113,
  \dodoi{10.1088/0004-637X/692/2/L113}

\end{thebibliography}
\bibliographystyle{aasjournal}



\end{document}